\documentclass{article}
\usepackage{graphicx}
\usepackage{url}
\usepackage{amsfonts}
\usepackage{amsmath}
\usepackage{amsthm}
\usepackage{booktabs}
\usepackage{subcaption}
\usepackage{multirow}
\usepackage{newfloat}
\usepackage{listings}
\usepackage[utf8]{inputenc}
\usepackage[T1]{fontenc}
\usepackage{hyperref}
\usepackage[ruled]{algorithm2e}

\begin{document}

\title{Diversity of Skills and Collective Intelligence in GitHub}

\author {
    Dorota Celińska-Kopczyńska  \\
    { \small Institute of Informatics, University of Warsaw}
}

%\keywords{open source; diversity; similarity; GitHub; social networks.}

\maketitle

\begin{abstract}
A common assumption suggests that individuals tend to work with others who are similar to them. However, studies on team working and ability of the group to solve complex problems highlight that diversity plays a critical role during collaboration, allowing for the diffusion of information. In this paper, we investigate the patterns behind the connections among GitHub users in Open Source communities. To this end, we use Social Network Analysis and Self-Organizing Maps as the similarity measure. Analysis of textual artifacts reveals the roles of those connections.
We find that diversity of skills plays an essential role in the creation of links among users who exchange information (e.g., in issues, comments, and following networks). The connections in networks related to actual coding are established among users with similar characteristics. Users who differ from the owner of the repository report bugs, problems and ask for help more often than the similar ones.
\end{abstract}

\section{Introduction}
Open Source is more than the type of a~license; it turned into a~social movement that integrates developers with common purposes within communities. The interactions within communities form structures -- networks.
In real-world networks, connections are rarely established or maintained because of purely random factors. Usually, one can find patterns describing connections (edges) among agents (nodes). They can be generalized under terms homophily, if the agents bond with others of a similar sort, or heterophily, if otherwise.

Extensive research has been conducted on the role of homophily on link establishment in various networks, both informal (e.g., family ties or friendship) and formal (organizational environments) \cite{birds}. There is a~consensus among researchers that homophily is a~crucial factor in fostering links in informal setup \cite{lazarsfeld, marsden, lin}; however, the results are ambiguous when it comes to organizational structures \cite{burt, louadi}. On the one hand, homophily leads to more effective communication because similarities among individuals reduce the coordination costs. Participants may assume that they possess a~common knowledge background. On the other hand, heterophilic connections give access to a~diversity of ideas because different agents process the same information packages differently. That, in turn, enhances possibilities for finding optimal solutions.

Research on Open Source often focuses on the organizational and motivational aspects of software development's major tasks \cite{lakhani_hippel}. Researchers tend to investigate the output in the form of contributed source code \cite{onboarding, vasilescu_1, lima, mcdonald, dabbish, tsay}. However, collaboration in the successful project requires not only exciting and captivating but mundane and necessary tasks, e.g., bug-fixing, bug-reporting as well. Interestingly, those aspects of collaboration are rarely examined. 

Studies of diversity among GitHub users are limited. Authors either investigate traditional assortativity or rich-club coefficients \cite{lima} or analyze the possible gender discrimination \cite{gender}, the impact of female participants on overall productivity \cite{vasilescu_1} or cultural (nationality) differences \cite{ortu}. Although Vasilescu et al. \cite{vasilescu_2} surveyed developers on the perceptions of diversity within GitHub teams, addressing the problem of tenure and experience diversity, there has been no generalized empirical study inferring similarity from software production-related characteristics. We want to fill that gap.

This article analyzes whether GitHub users establish homophilic or heterophilic connections and what roles in project development those connections play. We start with a~traditional one-dimensional Social Network Analysis of node degrees. We discuss assortativity measures and rich-club coefficients. Next, we conduct a multidimensional analysis of the characteristics of developers. We apply this analysis results to data on communication among developers to investigate the roles of the different clusters of links. We show that heterogeneity of community members enhances the task-solving abilities of the team. By team-working, participants who possess different informational assets can compensate for their shortcomings. This way, we provide empirical proof of the presence of collective intelligence phenomena in GitHub.

\section{Related work}
Collective intelligence is a~form of universal distributed intelligence arising from the collaboration and competition of many individuals. It measures the team's ability to solve complex tasks \cite{levy, surowiecki, salminen}. The different information packages are processed by individuals through social interactions, eventually leading to providing the solution to a~problem.

Open Source communities form meritocracies. Members of the community collectively solve problems related to projects development. Heylighen \cite{heylighen} describes the process of collective problem-solving by the aggregation of mental maps. Mental maps are individual constructions based on subjective preferences and experiences.
Each individual can explore their mental map to develop a proposition that creates part of the solution. Assuming that the individuals are similar, we may expect their mental maps to be similar, too. However, 
a group consisting of identical members does not benefit from synergy: it is equal to its best member, where all of them are perfect substitutes. Group members' synergy increases the chances of finding the optimal solution and increases the group's collective intelligence. Here, we notice that the possibilities for the Open Source community's success emerge from the underlying collective intelligence within teams.

Finding evidence for collective intelligence phenomena is closely related to research on group task-solving. No matter the organizational environment (informal or formal one), there are three generally accepted analytical frameworks regarding team construction: similarity attraction theory (SA) \cite{byrne}, social identity and social categorization theory (SIC) \cite{tajfel}, and information processing theory (IP) \cite{salancik}. SA postulates that people tend to work with others of similar characteristics in terms of values, beliefs, and attitudes to them (homophilic relationships, ``birds of a feather flock together''). According to SIC, teamwork occurs among members of a given (self-categorized) group. SA and SIC imply negative consequences of diversity on group performance.

On the contrary, IP favors heterophily. In certain conditions, large, diverse groups can achieve better results than any individual in the group. The wisdom of crowds effect is claimed to be based on diversity, independence, and aggregation \cite{surowiecki}. Diversity among people usually refers to differences in demographic, educational, and cultural backgrounds, which eventually leads to differences how people represent and solve problems.    

Collective intelligence focuses on the positive aspects of synergy, but the aggregation of knowledge is troublesome. There may be situations when collective efforts fail \cite{hayek-challenge}. Groups are prone to reputational or confirmation biases. When team members think similarly, they may stick to accepted and typical solutions, instead of finding better, non-conventional ones. In extreme situations, if one of them errs, they cannot notice it; instead, the error may amplify.  

\subsection{Team-work in GitHub}
Every GitHub's registered user has their site that integrates social media functionality directly with code management tools \cite{marlow}. Each user's profile contains public information about their biographical characteristics (e.g., the optional description of a location, employer, personal site, and e-mail address), the list of public repositories, and the number of followers as well as the number of people the user follows. Everyone, even unregistered users of the service, may browse the profiles and download the public repositories. The contribution to the project occurs by sending the patches to the author of the original project via e-mail or by submitting the pull requests -- the proposed changes to the projects that may be accepted or declined by the maintainers of the project \cite{celinska_euromed}. If the registered GitHub user wants to commit to the original repository, they have to fork the existing repository (create their copy of the source code). As a result, the user (known as a member) gets access to the complete source code. The members may introduce their commits (proposed bug fixes and changes to the source code) locally but are also allowed to synchronize their repositories with the state of the original repository.

By default, public repositories in GitHub should be considered Open Source; the founders decided to support the development of Open Source software providing free-of-charge hostage for such projects \cite{github-blog-11}.
Even if Open Source is a collaborative effort (so the team's synergies take place), much research focused on individual motivation \cite{crowston, david_shapiro, hann, hertel, meissonier, osterloh, roberts}. Generally, Open Source communities are instances of reputation-reciprocity environments. Reputation among developers’ peers is the measure of competitive success. Developers’ anticipated increase in reputation resulting from the number of fixed bugs encourages them to participate in team working on the Open Source code \cite{hars, brossi2006, raymond2, lerner}. Reputation gains do not motivate all developers equally -- Lakhani and von Hippel \cite{lakhani_hippel} showed that peer reputation stimulates ``gratifying'' technical tasks. However, it is not an incentive for mundane but necessary tasks, non-negligible in each software project. Furthermore, Lattemann and Stieglitz \cite{latteman} suggested that reputation inspires developers rather than bug fixers or managers. In GitHub, proxies of developers' reputation and ``social status'' emerge from the following network and the starring and watching networks. Popular and influential users have a~large number of followers interested in their coding activity, what projects they follow or work on \cite{herbsleb, dabbish, badashian}.

Developers expect future gifts in return. The reciprocity as a~motivational factor for developers was reported in various studies \cite{david_shapiro, bergquist, lakhani}. Lakhani and von Hippel \cite{lakhani_hippel} showed that reciprocity encouraged developers to perform mundane tasks. It is supposed that those who were offered help in the past are more eager to reciprocate after gaining more experience and knowledge \cite{vonkrogh}.
An extreme case is kinship amity \cite{fortes, zeitlyn}, according to which entity involves in altruistic behavior, working in favor of their relatives and close neighbors without expectation or requirement of the reciprocity. 

Contrary to individual motivations, studies on collective intelligence and diversity in GitHub or Open Source are limited. Daniel et al. \cite{daniel} investigated the relationships between diversity and community engagement and market success. They found that diversity in developers' reputation and role enhances market success and community engagement. Diversity of spoken language and nationality influences negatively community engagement but has a positive impact on market success. Vasilescu et al. \cite{vasilescu_1, vasilescu_2} suggested that increased gender and tenure diversity stimulates greater productivity when forming a software team. 

Since the results on team construction and the roles of different connections are ambiguous, we have decided to answer the following questions.

\textbf{\emph{RQ1. Do GitHub users exhibit homophily while establishing links? If so, what are the patterns of homophilic relationships?}}

SA or SIC suggest homophily; the kinship concept also implies it. Similarities should decrease the costs of communication. On the other hand, teams work for the ``greater'' good (a project), so the differences should be reconciled and exploited. That is why we want to analyze:

\textbf{\emph{RQ2. What are the roles of homogeneous and heterogeneous links in Open Source projects hosted in GitHub?}}

Finally, since SA and SIC suggest that work-group heterogeneity can lead to confusion, stress, and conflict \cite{horwitz} we want to find:

\textbf{\emph{RQ3. Does the increased diversity in teams affect the sentiment of conversation negatively?}}

\section{Methods and data}
We followed a mixed-methods approach. We started with a statistical description of the networks using traditional Social Network Analysis and multidimensional technique. To assist in interpreting the quantitative findings, we assembled data set on text artifacts, on which we modeled the roles of the connections among GitHub users in the project development.

\subsection{Data on users' characteristics and relationships} GitHub data is huge, and the service is continuously evolving. Therefore the complete download of the data is impossible \cite{perils, ght_firehose}. To minimize the number of missing observations, we use a data set combined from three sources: GHTorrent project \cite{ght_firehose} (dump dated Feb. 1, 2017), GitHub Archive project \cite{gha}, and our database obtained by web-scraping GitHub in 2016. Usually, using GHTorrent should be enough for most of the research activities concerning relatively big and popular projects \cite{perils}. Merging with GitHub Archive improves the data set's quality regarding smaller and lesser-known projects. Merging with scraped data and the API list of the registered users enables a more robust generalization of the results for the users. GHTorrent does not update the user table basing on the registrations, but on the public activity, mostly in the so-far collected projects. Since GHTorrent refreshes all users and all repositories every few months \cite{ght2}, it may take a reasonable time until it collects new account data. Crucial for our application, GHTorrent also offers limited possibilities for the qualitative analysis of the collaboration, e.g., text mining; the original text entries are usually not available.

The data about GitHub's registered users is publicly available but very distributed. To combine the data, we used a set of heuristics. GitHub Archive data in JSON format contained information about the unique id, the same as in web-scraping data. Data coming from GHTorrent lacked this information, but we were able to merge records using users' logins and the registration date. The standardized timestamps merged the events we used to create networks. We analyzed six types of social networks: forking, pull requests, issues, comments, following, and starring. We considered user accounts sharing the gravatar or using the same login at different times the same user. Thus such accounts were merged into a single observation using the Find-Union algorithm \cite{cormen}. The data we utilize and collect is publicly available. We do not process personal data, but even if we did, providing such data is not mandatory, the user decides -- that is why we suppose that the information we collect does not violate anyone's consent.

The construction of the final database consisted of two stages. In the first one,we combined the raw data sets. In the second one, the unification key was applied to the combined data sets to remove the repeated entries or combine the information for the computation of the final variables. GHTorrent became the basis of the merging. In the cases in which network information in three data sets was available, we merged the scraped data at the end. 

The population of this research consists of users of GitHub registered in 2007-2014, i.e., 10,361,315 entities. We limited the span of registrations to make sure the developers had a chance to gain popularity, learned how to use the service and had enough time to start collaborating with others. We also disregard the ``organization'' accounts due to the non-random missing data patterns related to those profiles.

\begin{table}[!bt] 
  \centering
  \caption{Number of entries in databases. Scraping, GitHub Archive, and GHTorrent report on the numbers of entries found before the merging process. After merging contains data after combining networks, users' unification, and limiting the span of registrations \label{number_obs}}
  \resizebox{0.98\linewidth}{!}{
    \begin{tabular}{l|c|c|c|c|c|c|c}
    \toprule
    Number of & \multicolumn{2}{c|}{Scraping} & \multicolumn{2}{c|}{GitHub Archive} & \multicolumn{2}{c|}{GHTorrent} & After merging \\ \cline{2-8}
    & Overall        & Unique        & Overall        & Unique        & Overall        & Unique      &  \\ 
    \midrule
    following relations    & 7,672,122      &  31,922       & 2,654,424      & 949,881       &  16,706,738    & 15,002,195  & 15,944,720 \\  
    forking relations       & 6,935,287      &  846,693      & 25,600,929     & 13,441,444    &  17,662,591    & 5,503,106   & 24,276,564 \\
    issues relations        & n/a            &  n/a          & 37,389,274     & 11,857,650    &  55,471,260    & 29,939,636  & 8,239,025  \\             
    pull requests relations & n/a            &  n/a          & 54,390,705     & 12,027,388    &  47,917,259    & 5,553,942   & 4,787,584  \\             
    starring relations      & 46,583,597     &  n/a          & 66,392,262     & 12,787,360    &  63,172,922    & 9,568,020   & 61,163,902 \\            
    commenting relations    & n/a            &  n/a          & 62,910,591     & 10,148,954    &  67,651,808    & 14,890,171  & 7,341,043  \\            
    identified logins (raw) & 10,938,114     &  4,278,933    & 7,729,951      & 177,612       &  16,373,567    & 6,587,227   & 8,158,116  \\            
    identified repositories & 10,745,523     &  2,911,806    & 47,502,004     & 34,019,546    &  10,121,067    & 2,301,991   & 29,636,285 \\ \hline
    Timespan covered & \multicolumn{2}{c|}{2007-2016} & \multicolumn{2}{c|}{2011-2017} & \multicolumn{2}{c|}{2007-2017} & 2007-2014 \\
    \bottomrule
  \end{tabular}}
\end{table}

Table \ref{number_obs} summarizes the merging processes. Columns ``Overall'' contain the number of entries found for the given vendor (no limitations). Columns ``Unique'' contain the number of entries found for the given vendor that did not repeat in other vendors (no limitations). The column ``After merging'' contains the number of entries after two steps of merging. The difference between the sum of unique entries for each vendor and the numbers in the ``After merging'' column comes from imposing limitations on the registration period and the users' unification. Identified logins do not equal the non-deleted users.

For clustering purposes, initially, we wanted to include all the users registered during 2008--2014, who did not delete their accounts (3,915,138 observations). However, the presence of users with no public activity significantly worsened the results of the clustering. Self-Organizing Maps (SOMs) tended to cluster only those observations correctly; thus, we removed this group of observations. We also removed the group with incidental activity (the sum of written issues, comments, and commits lower than 10). We standardized log-transformed variables to mitigate the impact of the remaining outliers and the range of the variables. For taking the logarithms, we incremented the values of the variables by five. Scaling by a~small factor (usually one) is a~generally accepted practice in such cases. The distributions of the variables were biased towards low values (0-10), so the standard treatment could overestimate the differences between them. E.g., for the values of our variables, the difference between zero, one, and two is negligible; the shift by one makes those differences non-linear. Shifting by five reduces this effect. The final sample contained 1,886,642 GitHub users. 

\subsection{Textual artifacts}
We analyzed the textual content created by GitHub users in the form of comments written in issues threads and bodies of the issues (the first comment in a~thread). First, we used Compact Language Detector 2 (CLD2) \cite{cld2} to assign the language to the comment. Secondly, we limited our sample to the items detected as English. Out of 26,508,924 comments retrieved, 22,830,296 were listed as English. Even if the comments in issues on GitHub fundamentally differ from the communication in other social media -- they aim at fostering the production of a~working product -- they are still generated via the Internet. We needed the preprocessing:
\begin{itemize}
\item Internet-specific content: we replaced emojis, emoticons, and slang with tokens containing their verbal description enclosed in colons;
\item source code was replaced with [code-snippet] token;
\item automated responses were replaced with [automated-message] token;
\item URLs for GitHub, Twitter, or StackOverflow were replaced with tokens [github $\vert$ twitter $\vert$ stackover-link];
\item @mentions (a typical way of notifying somebody of the post) were replaced with [user-mention] token, repository names with [repo-name] token;
\item each comment formed a~single line; if there were no [.?!] at its end, ``.'' was inserted;
\item we expanded popular English short forms.
\end{itemize}

For the needs of text mining, additional replacements were introduced. We merged negations (``not'') with the following word, a~standard treatment (removal of numbers, punctuation, stop words, and stemming) was also carried out.

The information about the similarity of the users was absent in the textual data. To combat this, we merged the textual data with the results of the clustering. The sample was limited to the comments for which we could
infer the distance (similarity measure) between the interlocutors. Information on the sequence of the comments was not available. We also had little possibilities to find who is the addressee of the message. That is why we only analyzed the comments in which the commenter differs from the repository owner. The final sample consisted of 5,647,858 comments. 

The distance between two users is the distance between the neurons they are mapped to by the SOM algorithm and serves as the similarity measure (the higher the distance, the less similar users are). We took the rounded mean over the results of multiple runs of the SOM algorithm. Then, we grouped the comments based on the inferred distance. We did not find any comments for distances from 11 to 13. There are two possible explanations: our data set contains natural missing observations that are impossible to impute (e.g., periods in which GitHub API does not return the textual data); a~part of those comments was written in languages different than English. Table \ref{comments_data} summarizes how many items we found in the given distance.

\begin{table}[!bt]
  \centering
  \caption{Number of textual items found for users in the given distance. The increasing distance measures increasing dissimilarity \label{comments_data}}
  %\resizebox{\columnwidth}{!}{
\begin{small}
\begin{tabular}{lcccc} \hline
\toprule
Distance     & Comments & Bodies  & Sentences (comments) & Sentences (bodies) \\
\midrule
0  & 75,271  & 4,072   & 165,486   & 11,037  \\ 
1  & 354,991 & 25,177  & 1,074,645 & 124,342 \\ 
2  & 518,739 & 45,140  & 1,241,702 & 133,713 \\ 
3  & 483,693 & 48,746  & 1,192,105 & 144,672 \\ 
4  & 470,559 & 53,782  & 1,171,277 & 172,917 \\ 
5  & 435,391 & 51,666  & 1,074,032 & 164,828 \\ 
6  & 431,251 & 50,457  & 1,116,885 & 159,916 \\ 
7  & 632,704 & 62,508  & 1,475,491 & 206,722 \\ 
8  & 689,399 & 104,259 & 1,808,844 & 355,135 \\ 
9  & 949,094 & 149,407 & 2,743,520 & 560,525 \\ 
10 & 76,927  & 11,552  & 255,780   & 52,574  \\
\bottomrule
\end{tabular}%}
\end{small}
\end{table}

\begin{figure}[tb]
  \begin{flushleft}
    \includegraphics[width=0.45\linewidth]{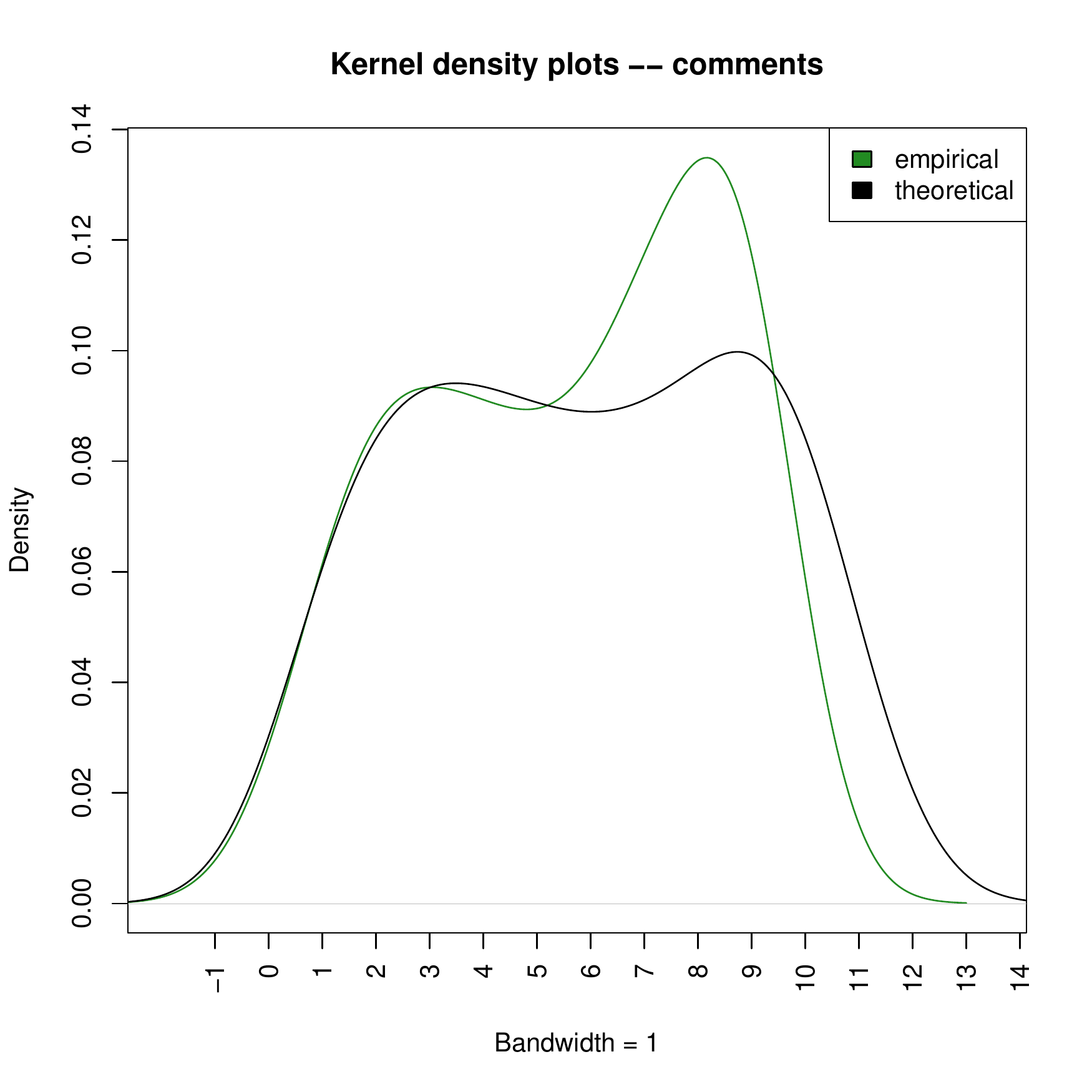}
    \includegraphics[width=0.45\linewidth]{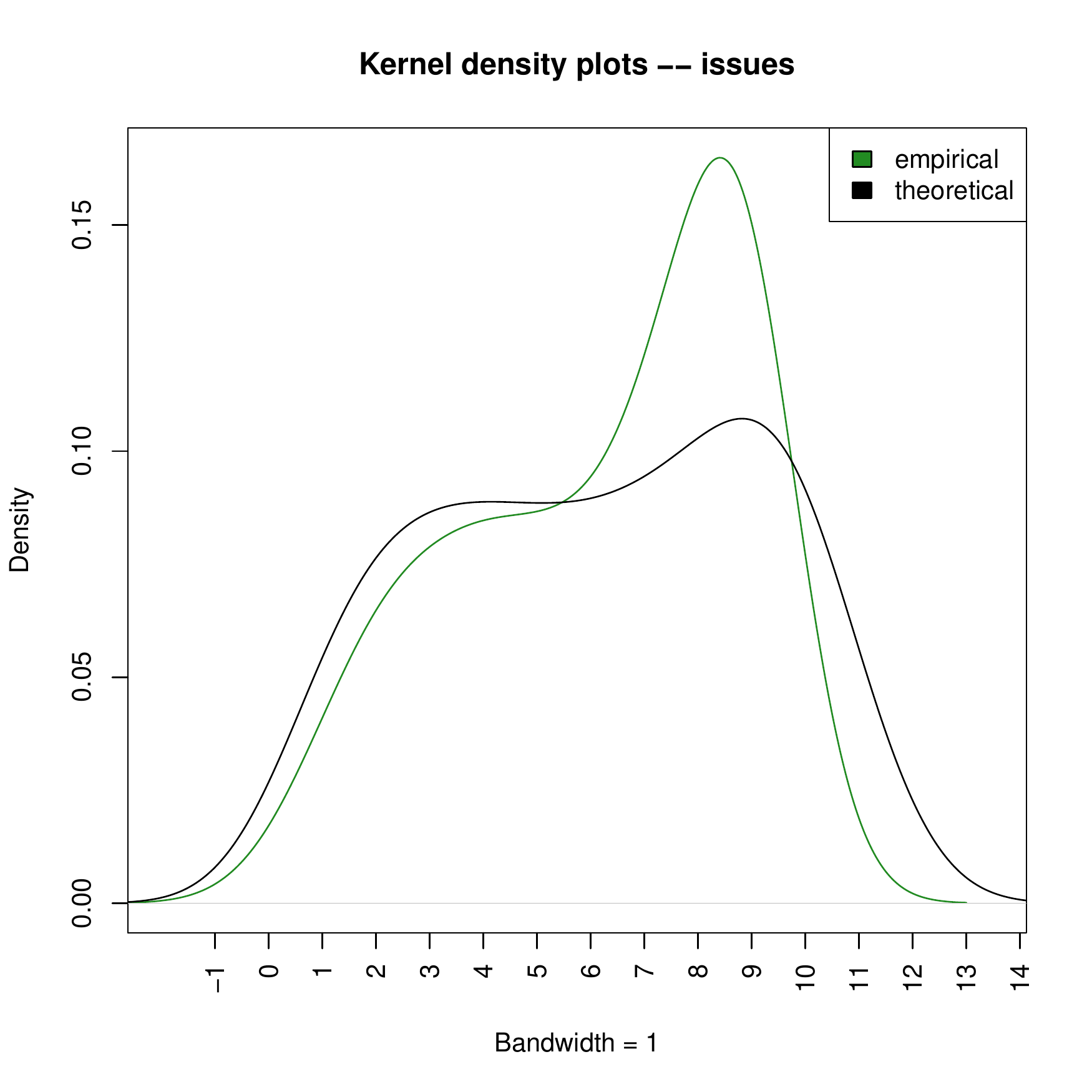}
    \caption{Theoretical and empirical density for number of comments and issues (bodies) for users of increasing diversity \label{comments_dens}}
\end{flushleft}
\end{figure}

Even if we miss data for distances 11-13, we do not find it threatening. Figure \ref{comments_dens} depicts the comparison between the theoretical distribution of issues or comments (based on the leave-one-out simulation for the means of the results) and the empirical one. Out sample is characterized by an overpopulation of the textual content created by dissimilar users. However, the shapes of the density functions follow the same pattern, so we assume our sample is representative.

\subsection{Modeling tendencies behind links}
Measuring tendencies among nodes to bond with similar or not is a~well-known topic in theoretical research on social networks. The traditional way of measuring homophily means computing correlation coefficients for node degrees. For multi-dimensional modeling, the family of Exponential Random Graph Models (ERGM) can be used. This approach is not valid in our case. First, ERGMs are computationally complex \cite{ergms}. Second, the networks' structures are intercorrelated \cite{celinska_euromed, celinska_smes}, so the error terms among the individual equations would be correlated, leading to biased results. We need a~simpler approach that would be robust to interdependence among the variables and easy to implement. For this reason, we use Self-Organizing Maps to measure similarity in a~multidimensional way.
 
\subsection{One-dimensional analysis}
Assortativity represents to what extent nodes in a~given network associate with other nodes in this network, being or not of a~similar sort \cite{newman}. We express assortativity as a~scalar value in the range [-1,1]. 1~denotes an entirely assortative network, i.e., the one in which the nodes with low degree link to other low-degree nodes. On the contrary, -1 denotes completely disassortative network, in which the low degree nodes connect to high degree ones. 

For directed networks, we analyze three different types of assortativity \cite{noldus}. $r_{in}$ denotes in-degree assortativity, $r_{out}$ is out-degree assortativity, and $r_d$ measures overall assortativity. We diversify the assortativity for directed networks because nodes may exhibit different tendencies when considering the connections' direction. Assume that node $v_1$ has $l$ out-going links and 0 in-going ones and node $v_2$ has 0 out-going links and $l$ in-going ones. The degree for both nodes is the same and equals $l$. If connected, these nodes are assortative in degrees, however, disassortative in both in- and out- degrees.

The networks in our study are scale-free with non-normal degree distributions (see, e.g., \cite{lima, dabbish} for following and collaboration networks, \cite{allaho} for power-law in issues and \cite{chatziasimidis, jarczyk} reporting on power-law behavior in the starring/watching network), which threatens the standard computation of assortativity (Pearson coefficient). To combat this, we discuss both the Pearson and the Spearman coefficients \cite{litvak}. To minimize the impact of the outliers, we calculate the logarithms of values (shifted by one, since zeros may occur).

Rich-club coefficient inspects the tendency of high degree nodes to form tightly interconnected communities \cite{zhou_richclub}. Whereas assortativity quantifies two-point correlations and accounts for quasi-local properties of the nodes in the network, the rich-club coefficients are computed as global features within a~restricted subset. They allow for a closer inspection of the structure of homophily. Generally, the two techniques' results can be similar, but they do not have to be.

Rich-club coefficient measures how many edges are present between nodes of degree at least $k$ in comparison to how many edges there could be between these nodes in a complete graph (concerning the direction of the link). Since the coefficients' values increase monotonically with an increasing degree even for random networks \cite{colizza}, we need to apply a normalization method. The coefficient's value computed from the empirical network is scaled with the coefficient's value computed from the maximally randomized network with the same degree distribution.

\subsection{Multidimensional similarities}\label{sec:kohintro}

Discussing similarity in data is straightforward if we compare at most two dimensions.
However, if we are interested in possible non-linear relationships or want to introduce more variables
simultaneously, the analysis becomes challenging. We want to reduce the
dimensionality in data without significant loss of information. Two popular approaches include Principal Component Analysis (PCA) and Self-Organizing Maps (SOMs). In comparison to well-known multidimensional techniques of reducing dimensionality, SOMs have many advantages. They do not impose any assumptions on the variables' distributions and do not require independence among variables. They allow for solving non-linear problems. At the same time, they are relatively easy to implement and modify \cite{kohonen, soms}.

The SOM algorithm's core idea is using a~deformable template (a manifold) to translate data similarities into spatial relationships. For our application, SOMs' crucial feature is that the resulting spatial locations have interpretation. Two stages of SOM's training: competition and adaptation, force the neighborhood to attract similar objects. This way, the distance between the neurons becomes a~measure for similarity: the further, the less similar objects are.

\section{Results}
\subsection{Homophily and heterophily in GitHub -- general tendencies}

In Table~\ref{hetero_assortativity} we provide the computed assortativity measures for the six networks considered in our study.
\begin{table}[tb]
  \centering
  \caption{The assortativity of networks analyzed in the study \label{hetero_assortativity}}
  \begin{tabular}{lcccccc}
  \toprule
  Network                     & \multicolumn{2}{c}{Deg. Assortativity} & \multicolumn{2}{c}{In-Deg. Assortativity} & \multicolumn{2}{c}{Out-Deg. Assortativity} \\ \cline{2-7}
  & $r_d$  & $\rho_d$ &  $r_{in}$  & $\rho_{in}$ &  $r_{out}$  & $\rho_{out}$ \\
  \midrule
  following    & -0.127 & -0.009 & -0.104 & -0.044 & 0.030  & 0.098  \\
  starring     & 0.006  & -0.024 & -0.016 & -0.027 & 0.082  & 0.071  \\
  forking      & -0.036 & -0.012 & -0.051 & -0.061 & 0.110  & 0.124  \\
  issues       & -0.476 & -0.268 & 0.154  & 0.133  & -0.144 & -0.040 \\
  pull requests & 0.071  & 0.146  & 0.067  & 0.073  & 0.164  & 0.163  \\
  comments     & 0.033  & 0.040  & -0.004 & -0.015 & 0.138  & 0.122  \\
  \bottomrule
\end{tabular}
\end{table}
The Pearson and Spearman coefficients' values are similar (the only exception is the starring network, where the signs are opposite). In most cases, the networks are weakly disassortative in degrees (nodes with lower number of overall connections link more frequently to nodes with a higher number of overall connections). We observe usually weak disassortativity in in-degrees (nodes with a~high number of incoming connections on average tend to attract nodes with a low number of incoming connections). Finally, networks are assortative in out-degrees (nodes highly connecting with other nodes on average link to other nodes behaving in the same way). An interesting exception is the pull requests network, which is weakly assortative no matter which direction we analyze. That may suggest a~kind of sorting behavior while creating an edge in this network, and nodes exhibiting homophily.

\begin{figure}[!tb]
  \begin{center}
  \includegraphics[width=0.15\linewidth]{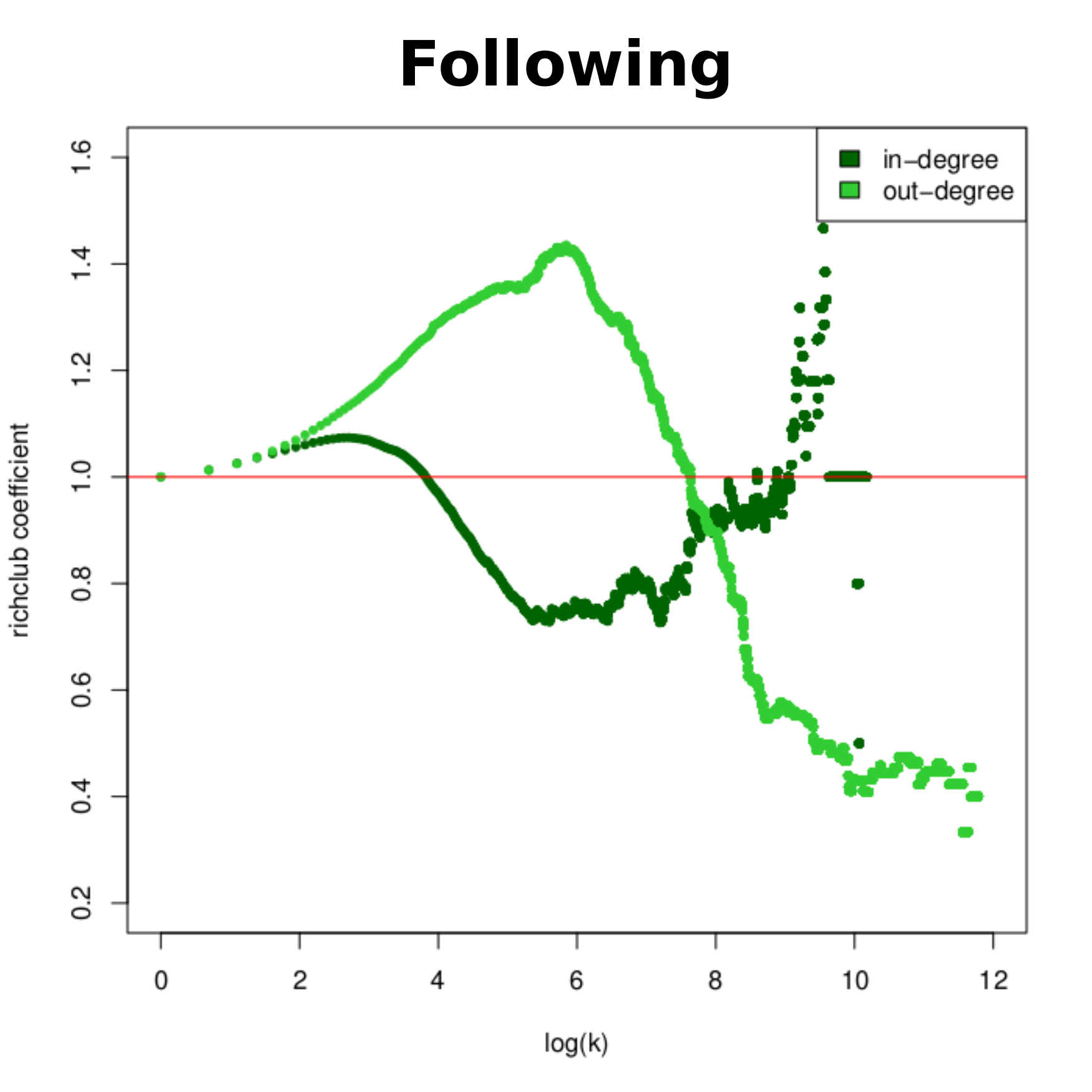} 
  \includegraphics[width=0.15\linewidth]{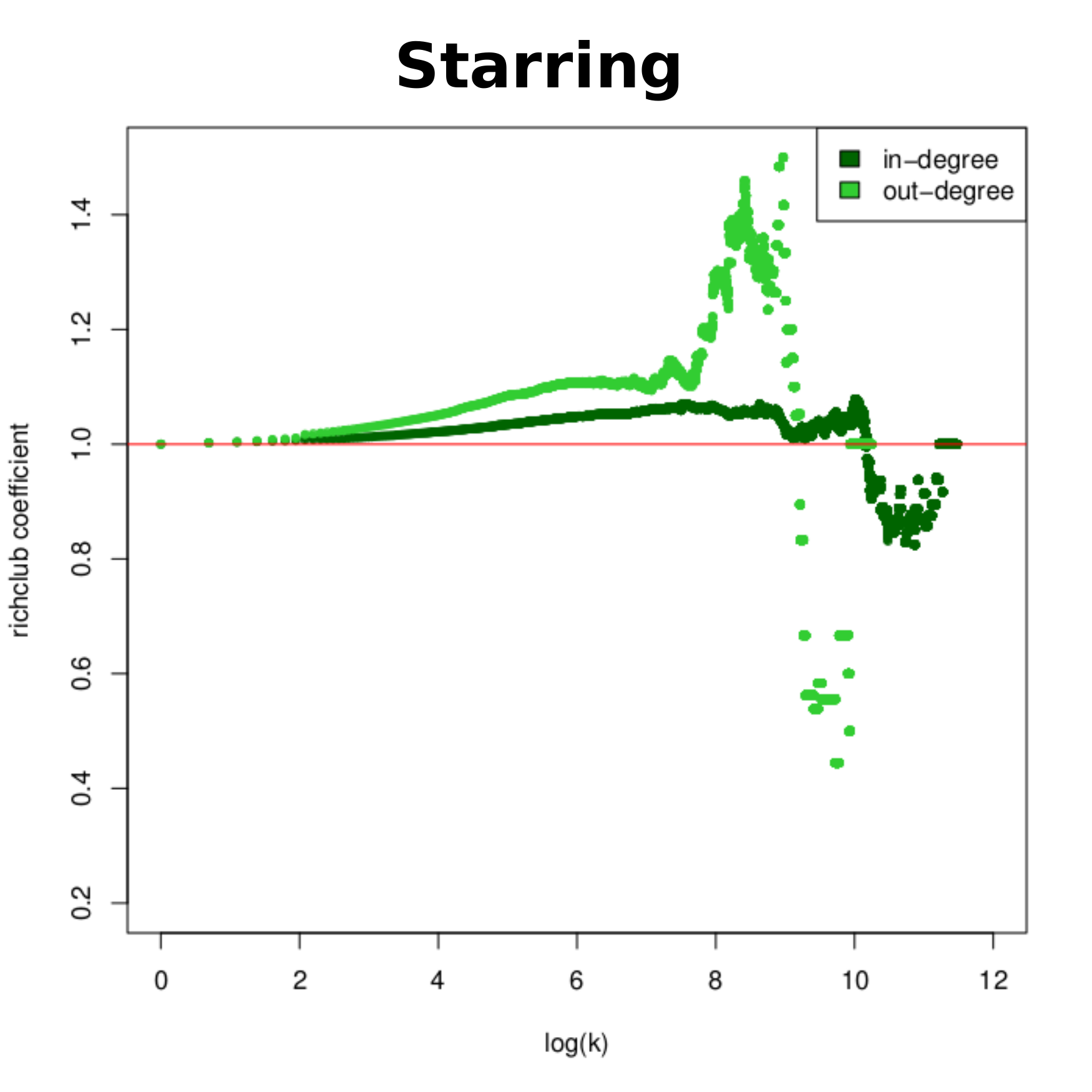}
  \includegraphics[width=0.15\linewidth]{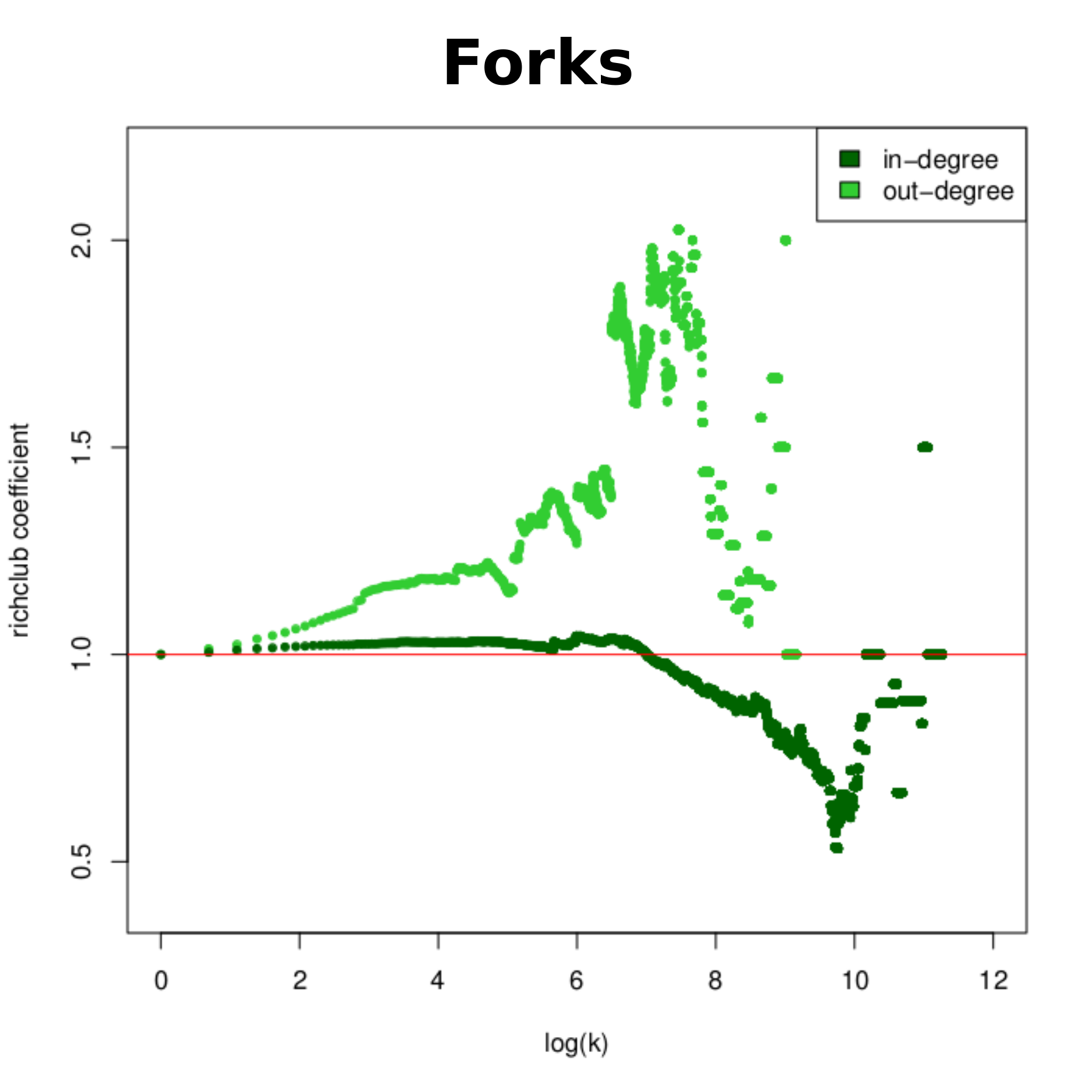} 
  \includegraphics[width=0.15\linewidth]{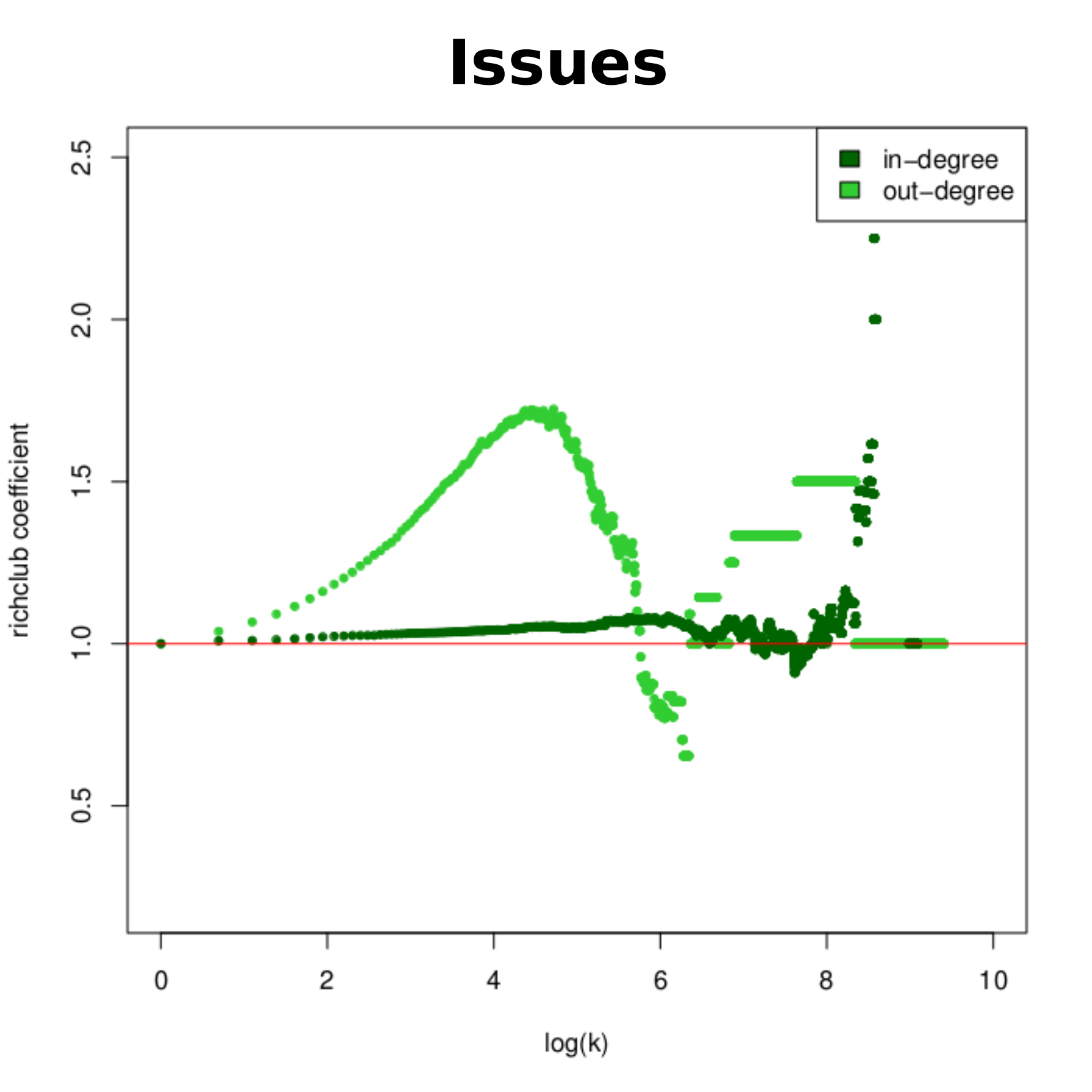} 
  \includegraphics[width=0.15\linewidth]{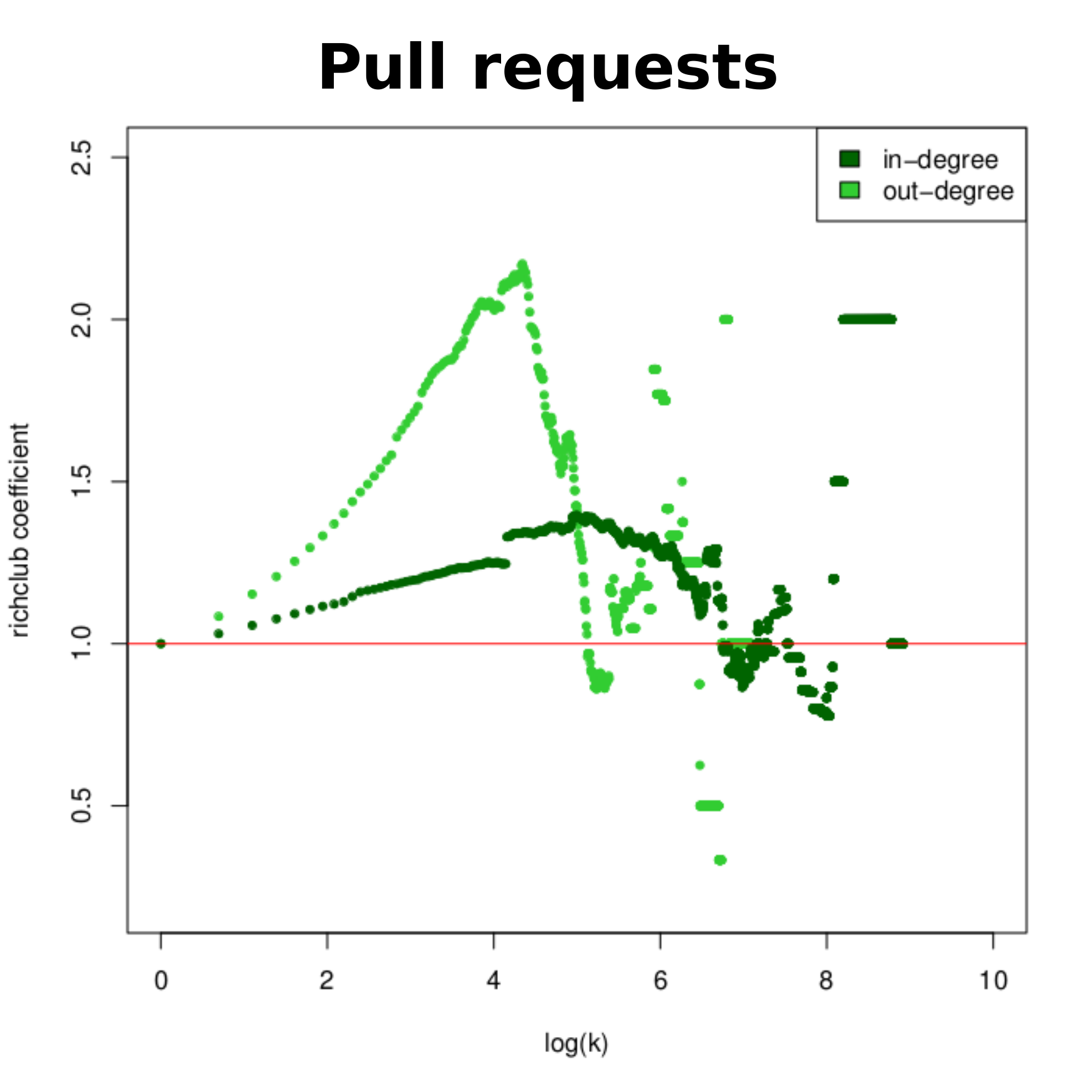} 
  \includegraphics[width=0.15\linewidth]{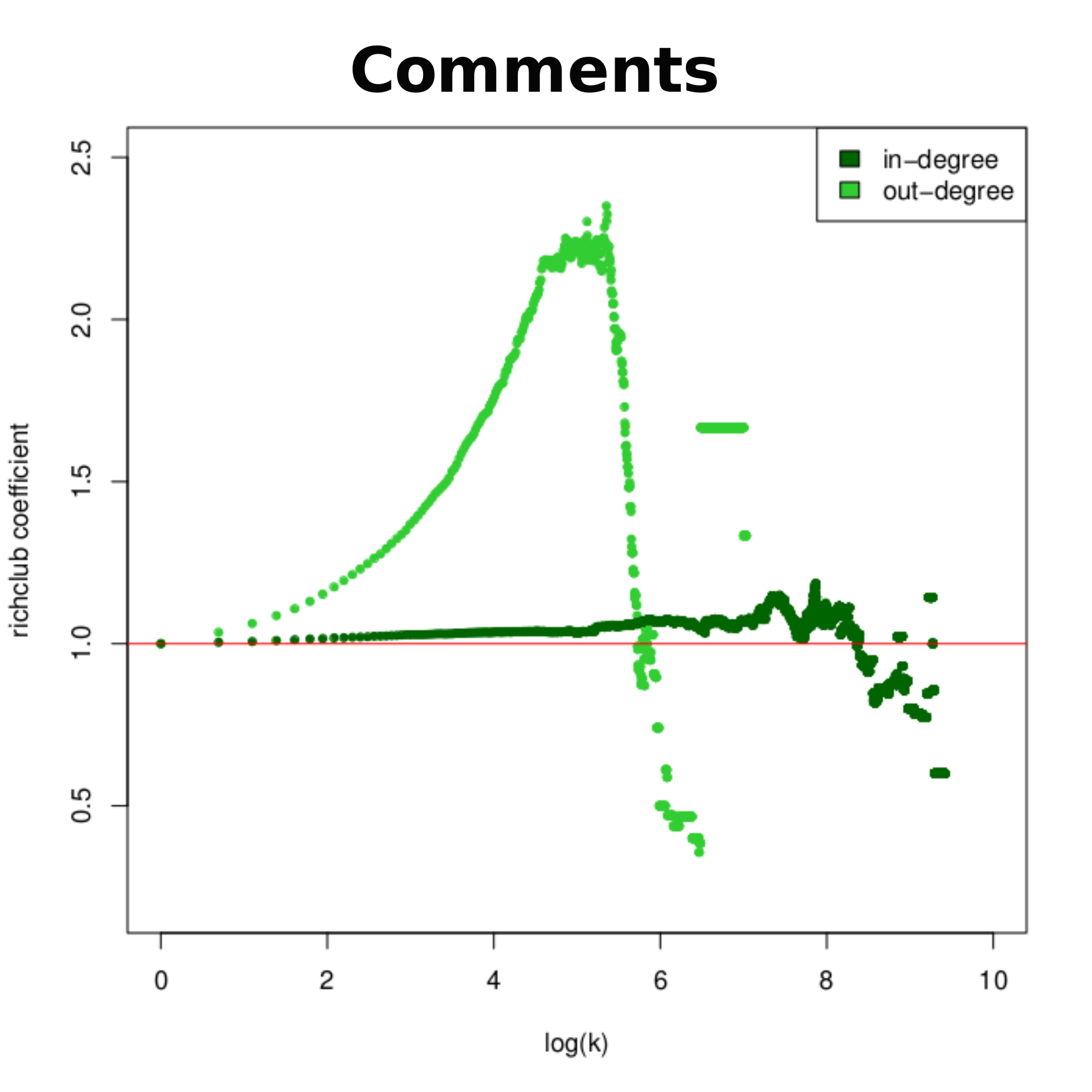}\\
  \includegraphics[width=0.15\linewidth]{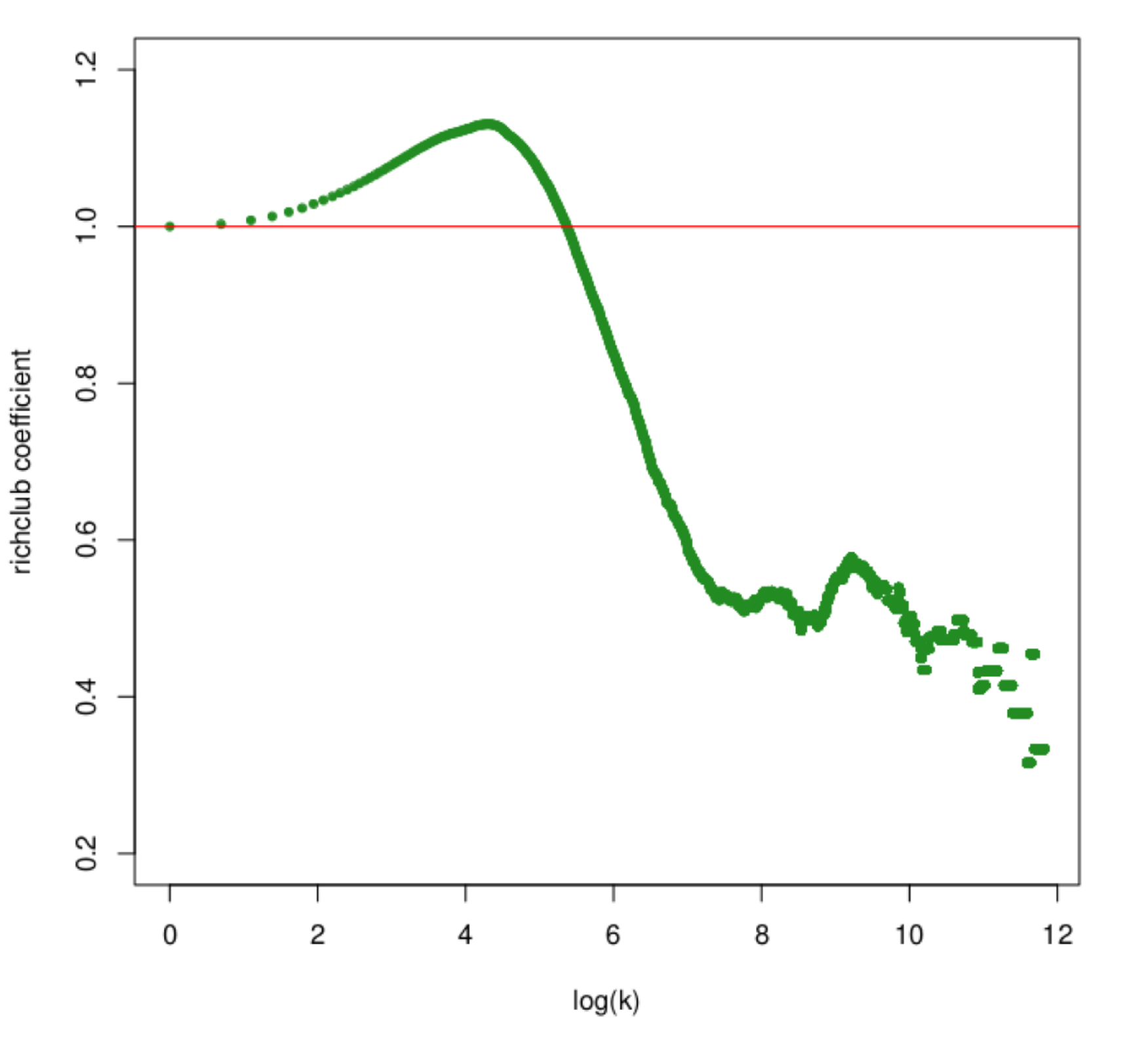} 
  \includegraphics[width=0.15\linewidth]{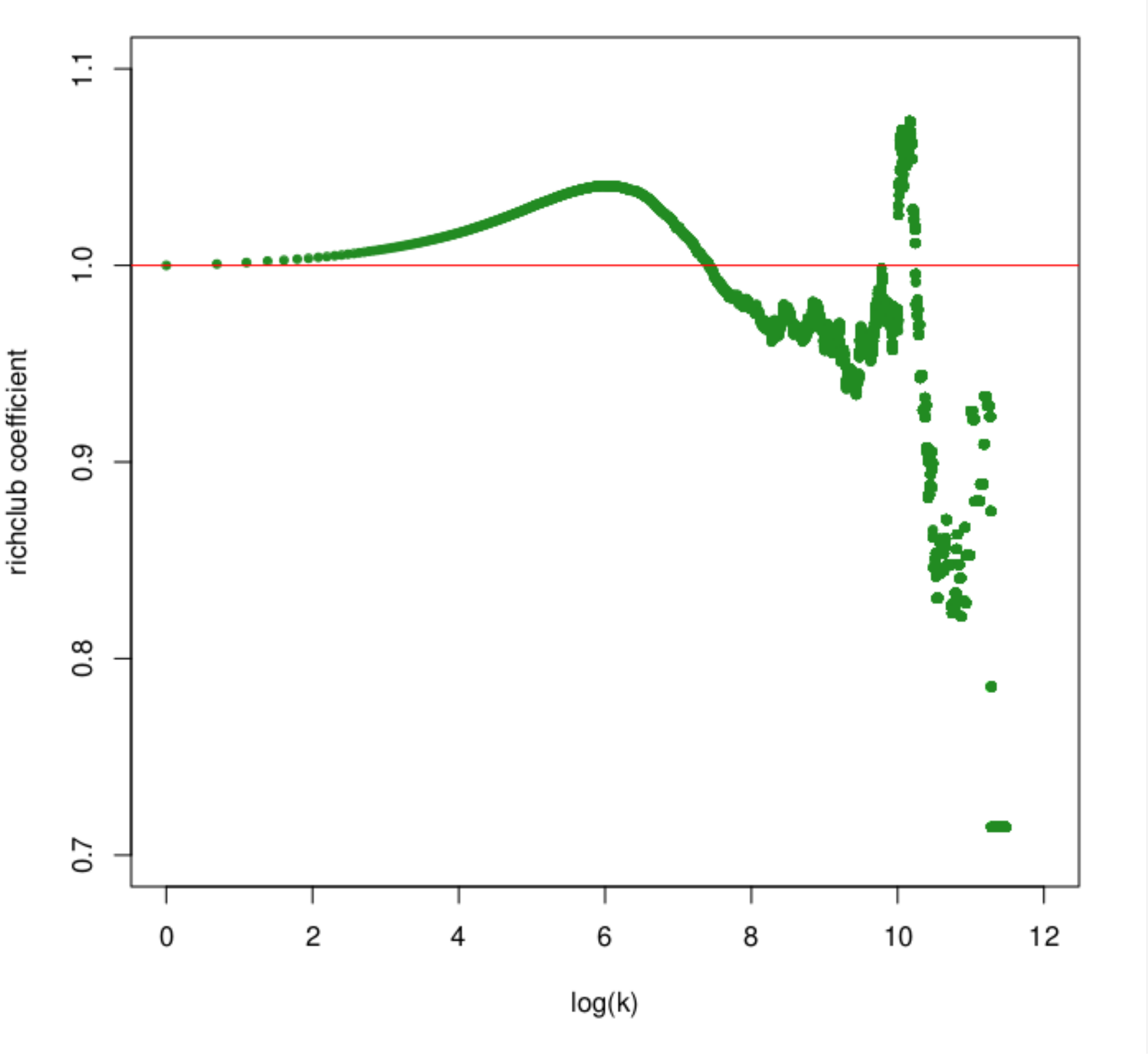}
  \includegraphics[width=0.15\linewidth]{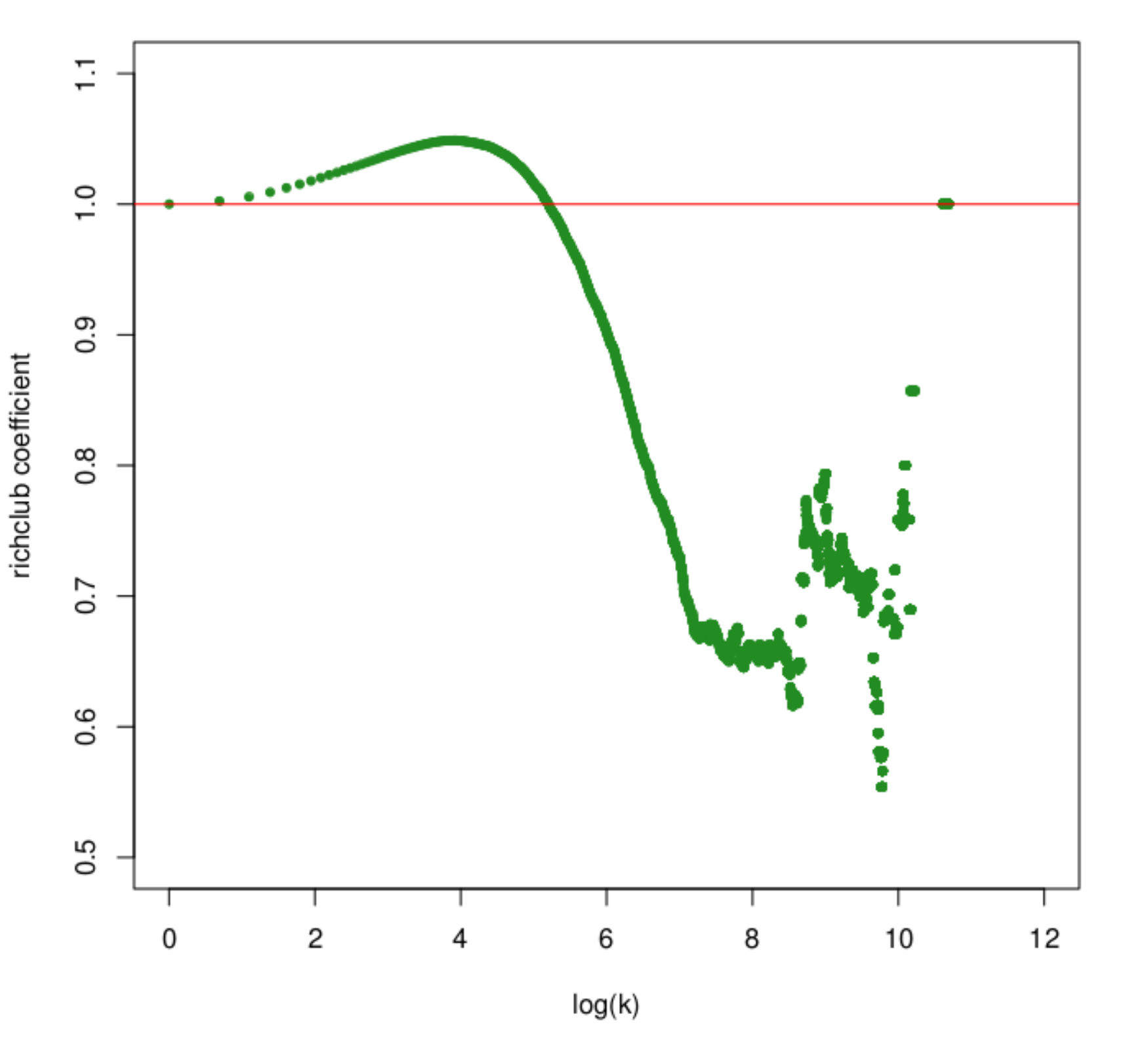} 
  \includegraphics[width=0.15\linewidth]{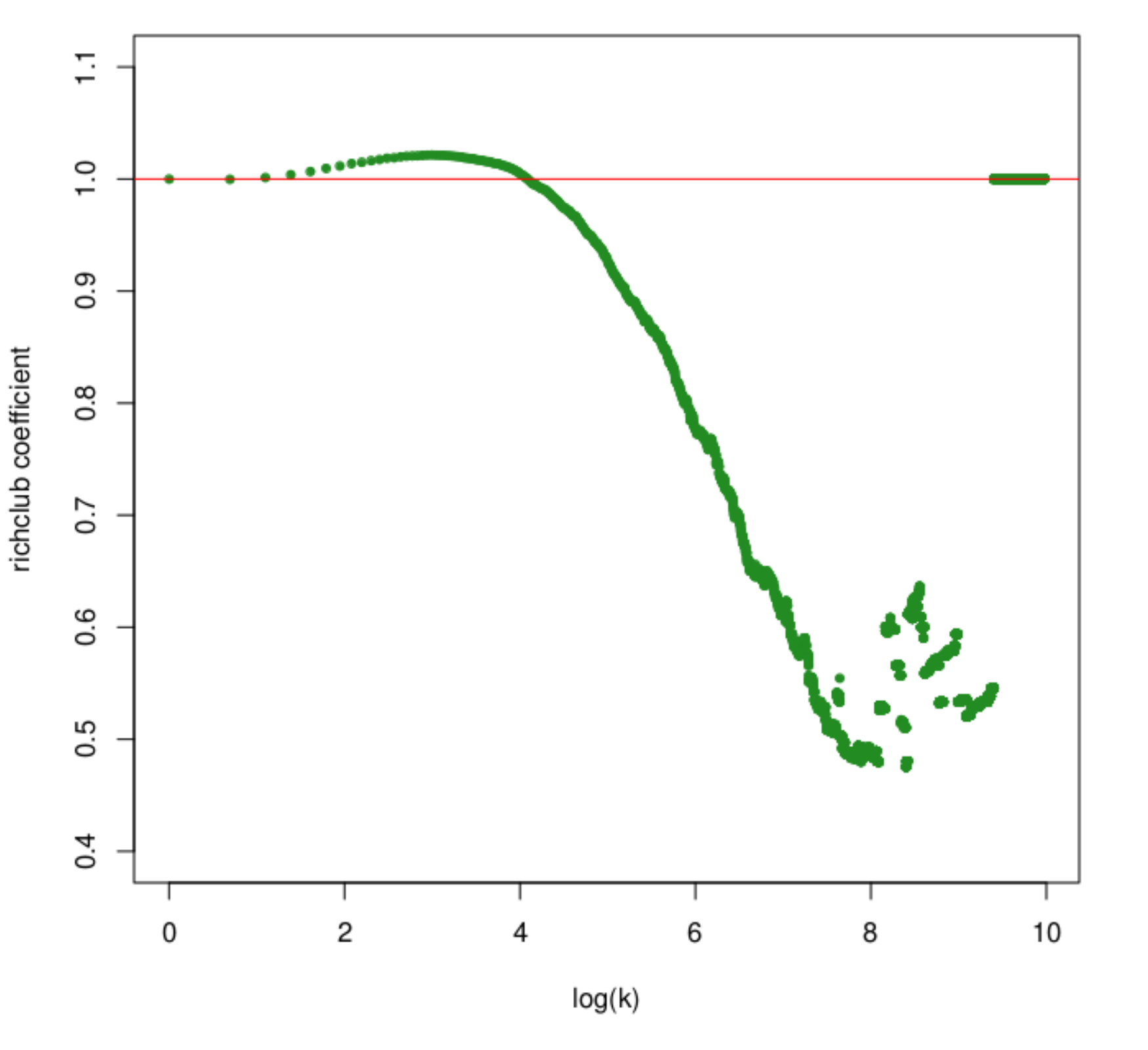} 
  \includegraphics[width=0.15\linewidth]{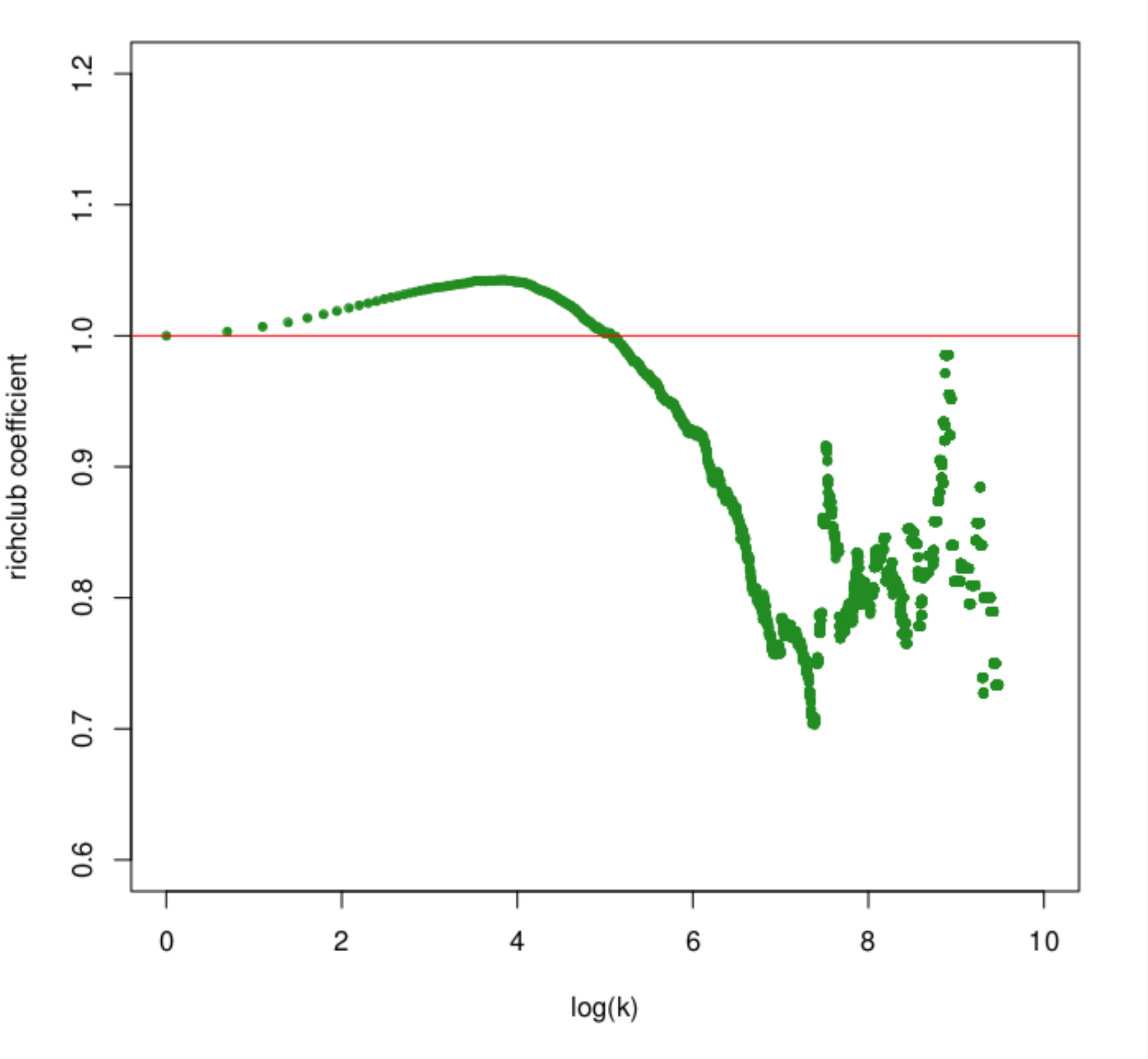} 
  \includegraphics[width=0.15\linewidth]{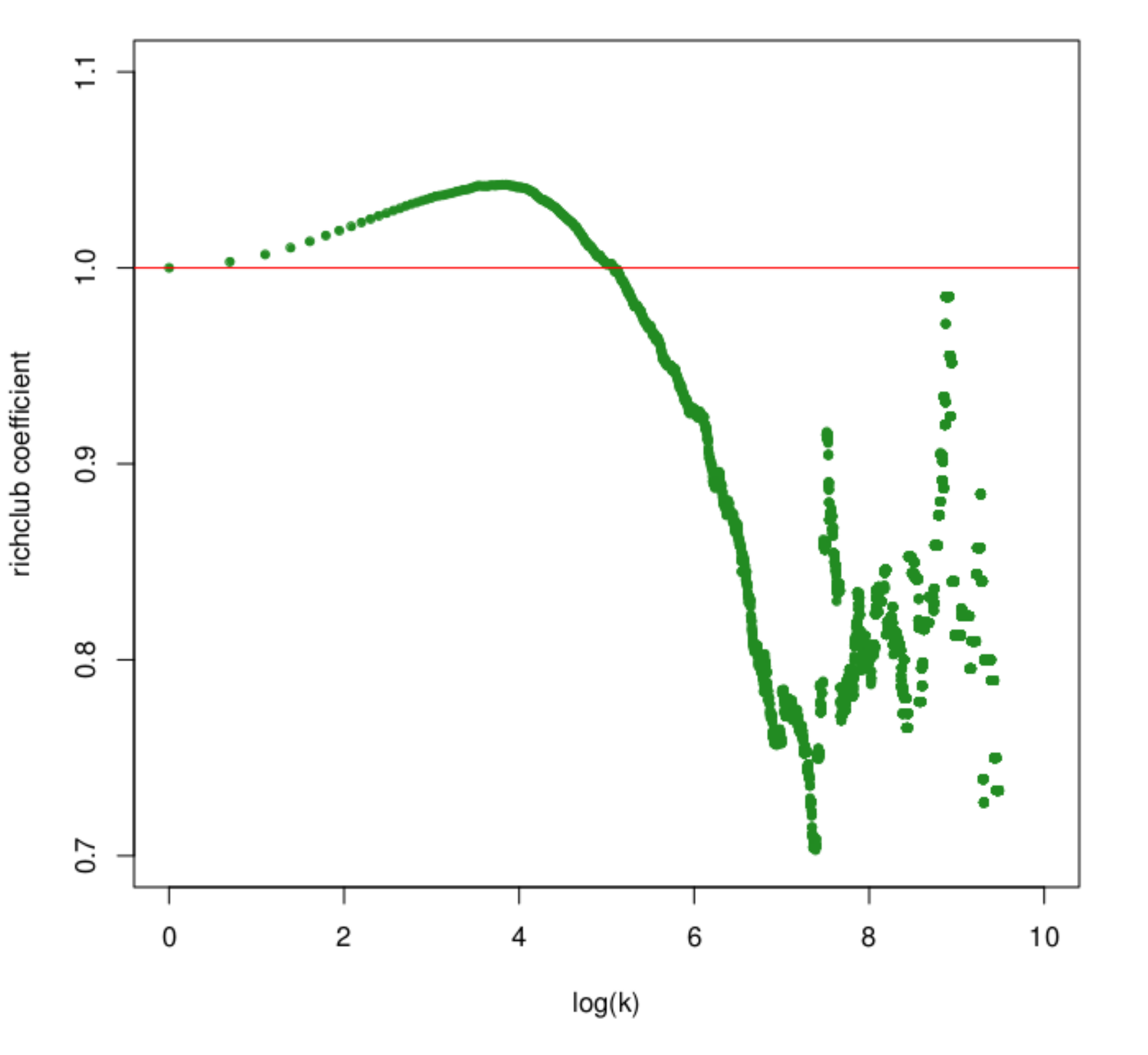} 
  \caption{Normalized rich-club coefficients as the functions of node degrees with (row 1) and without (row 2) respect to the directionality of the relations. The horizontal line represents the value of rich-club on a~maximally random graph. Values above (below) this line correspond to the presence (absence) of rich-club phenomenon with respect to the random case
 \label{richclub-directed}}
\end{center}
\end{figure}

Suppose do not take directionality into account. In that case, all networks show similar behaviors: one can observe forming rich-clubs among the nodes with relatively low degrees and the absence of rich-clubs among the top nodes (Figure \ref{richclub-directed}). Analysis of the coefficients' values suggests that popular users of GitHub tend to have connections with lower degree nodes rather than being tightly interconnected among similar nodes or nodes with even higher degrees. Again, an interesting case is the pull requests network. It is weakly assortative (i.e., generally exhibits homophily), but at the same time, the connections are not tight. The plot of the rich-club coefficient function of node degrees in pull requests network resembles a~general plot of a~more complex polynomial, showing changes in behavior concerning the degree; for most of the values of degrees indicating lack of rich-club phenomenon. Insights on the remaining networks are consistent with the conclusions derived from the analysis of assortativity, even if, in general, this is not mandatory. In theory, there can be even networks that exhibit both disassortativity and rich-club behavior.

While controlling for directionality in relations, one observes different patterns among nodes (Figure \ref{richclub-directed}). We observe rich-club phenomena for even quite high out-degrees (up to 1000, and for the starring network: 8000) in all considered networks. However, the patterns for in-degrees differ among the networks; users with moderately high in-degrees in the following network do not present the rich-club behavior; the changes in the trend are more observable compared to issues  or starring  networks. We notice an interesting difference in the patterns of rich-club phenomena while comparing pull requests network  with the following network: even if the trends for out-degrees are similar, they are different while controlling for in-degrees. 

\subsection{Clustering of GitHub users}
We present and interpret the results of applying a Self-Organizing Map utilizing Klein quartic to our data set. Following the literature, we utilized variables from five groups: 

\begin{itemize}
\item \textbf{Reputation}: the number of developer's \emph{followers}, \emph{stars obtained} by a~developer (the sum of stars given to developer's repositories), the eigenvector centrality of a developer  -- how important is the developer for the network, the number of times developers was \emph{forked\_by} others;
\item \textbf{Reciprocity}: the number of developers \emph{followed} by a~developer, the number of repositories \emph{forked} from others, \emph{stars given} to other developers' repositories, and \emph{commits} pushed to repositories owned by other developers;
\item \textbf{Communication}: the number of \emph{comments} written in own and others' repositories, and the number of \emph{issues} opened in own and others' repositories;
\item \textbf{Standardization}: the number of languages used in the developer's repositories, the fraction of specific usage intended languages in all of the languages used by the developer (specialization -- web-development, functional or scientific-intended languages);
\item \textbf{Information}: the number of \emph{repositories}, the \emph{year} of developer's registration, the number of \emph{commits} in the base repositories (the ones that were not forked) and in forked repositories.
\end{itemize}

\begin{figure}[bt]
\centering 
\includegraphics[width=0.75\linewidth]{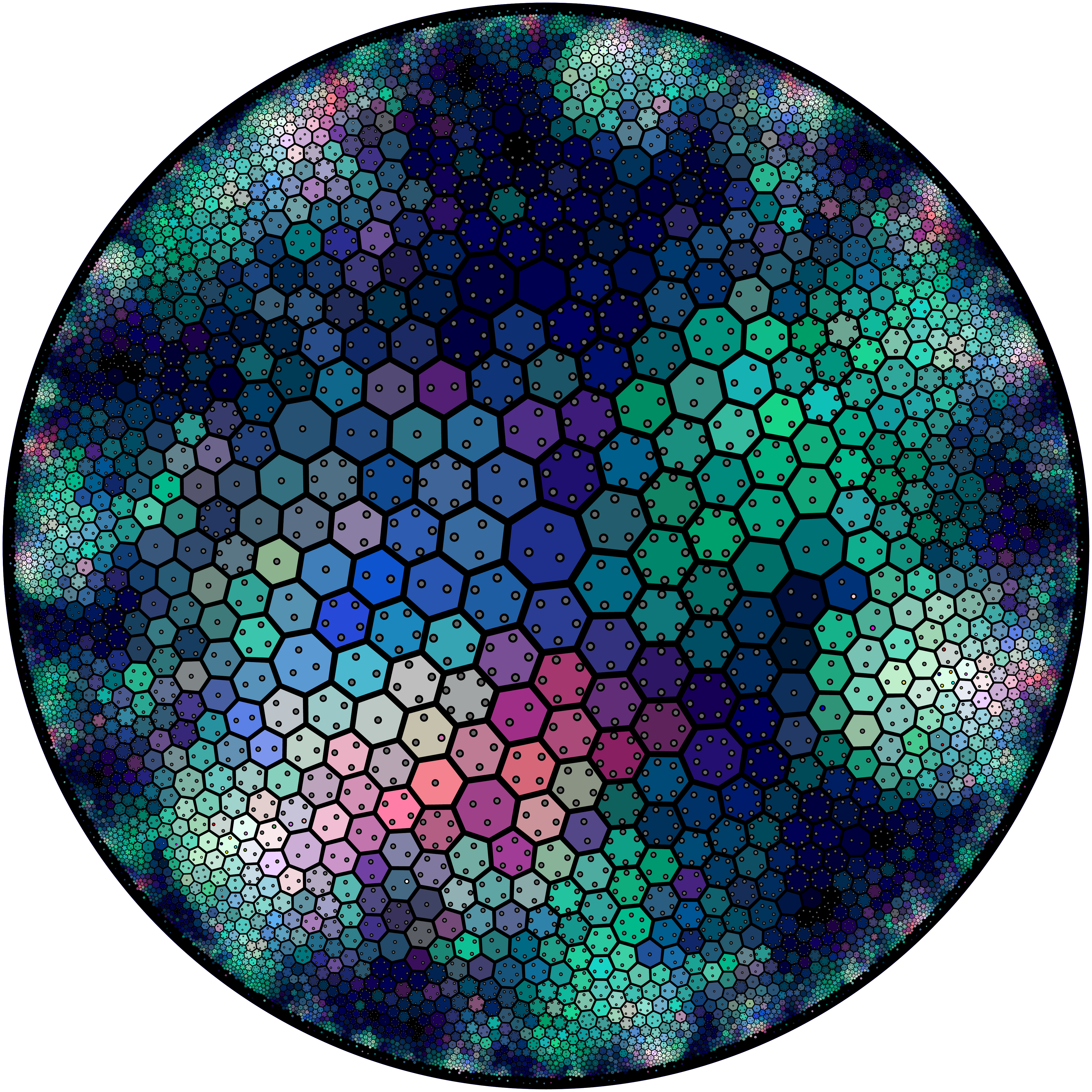}
\caption{SOM on Klein quartic: tiles colored with  number of repositories (R), commits by developer in their base projects (G), commits by developer in repositories they forked (B) \label{kohonen}}
\end{figure}

Figure \ref{kohonen} depicts the visual example result of the SOM clustering. Due to the exponential growth of the hyperbolic plane, we can fit similar objects closer to each other than possible in the Euclidean plane. We may see the smooth and gradual changes in colors; similar colors indicate similar characteristics. The darker the shade, the higher value of the given variable for the representative neuron. With the RGB additive color model, we can control up to three aspects at the same time. E.g., the light sea-green and cyan area in Figure \ref{kohonen} denotes the cluster of users who do not have many repositories (low values of red). However, they are very active in them: they frequently commit to their repositories (high values of green) and work with the source code of forks (medium values of blue). The users who belong to purple and pink areas are also active but characterized by a~different pattern. They have many repositories (high values of red), they rarely commit to their projects (low values of green), but often work on forked repositories (high values of blue). The brightest area highlights the extremely active developers.  The dots on the tiles approximate for the number of observations which were assigned to a~neuron (1: 15--1315; 2: 1334--1923; 3: 1935 --2372; 4: 2373--2793; 5: 2801--3227; 6: 3230--4099; 7: 4123--7214; 8: 7261--30130).

\begin{figure}[!tb]   
\begin{tabular}{ccc}
\includegraphics[width = 0.25\linewidth]{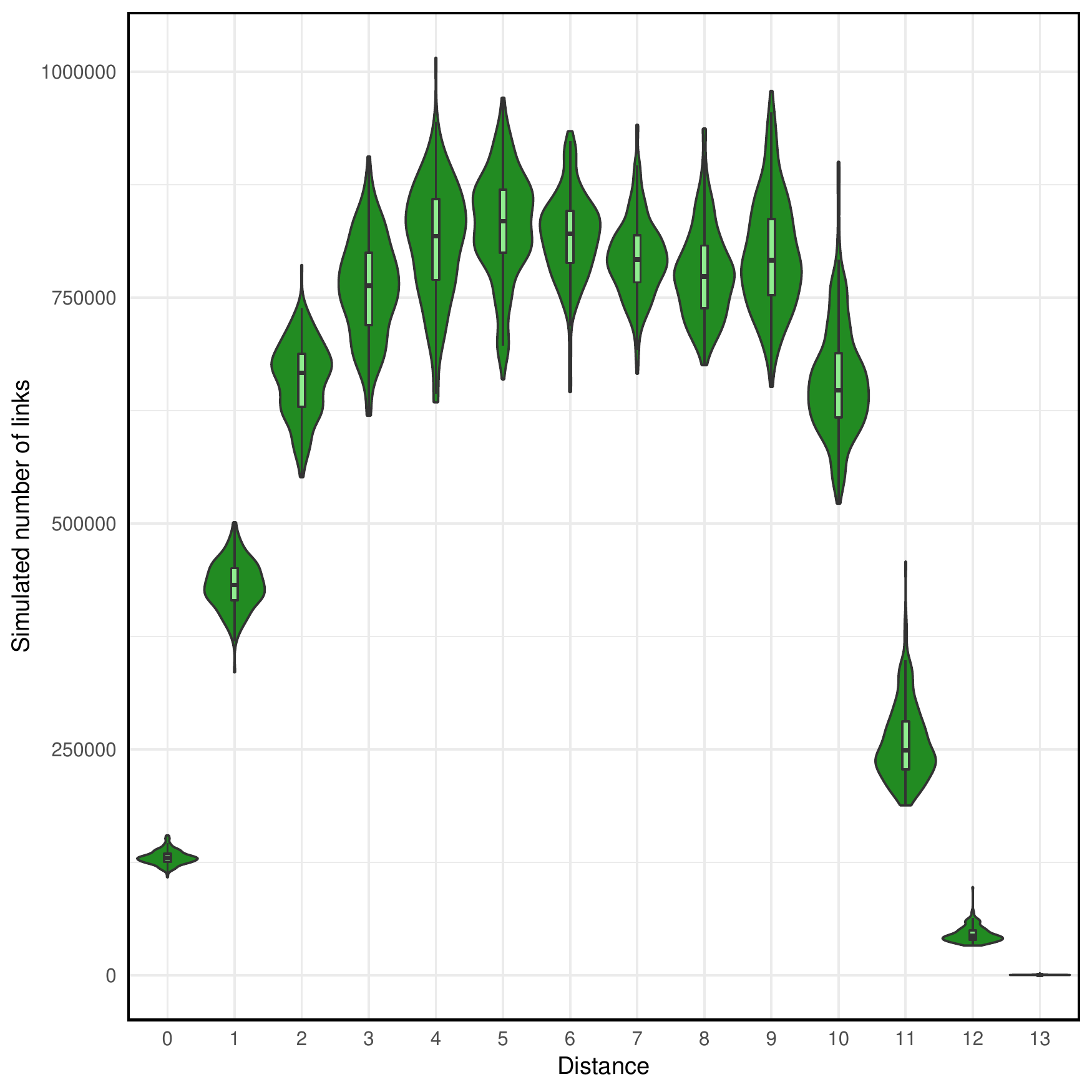} &
\includegraphics[width = 0.25\linewidth]{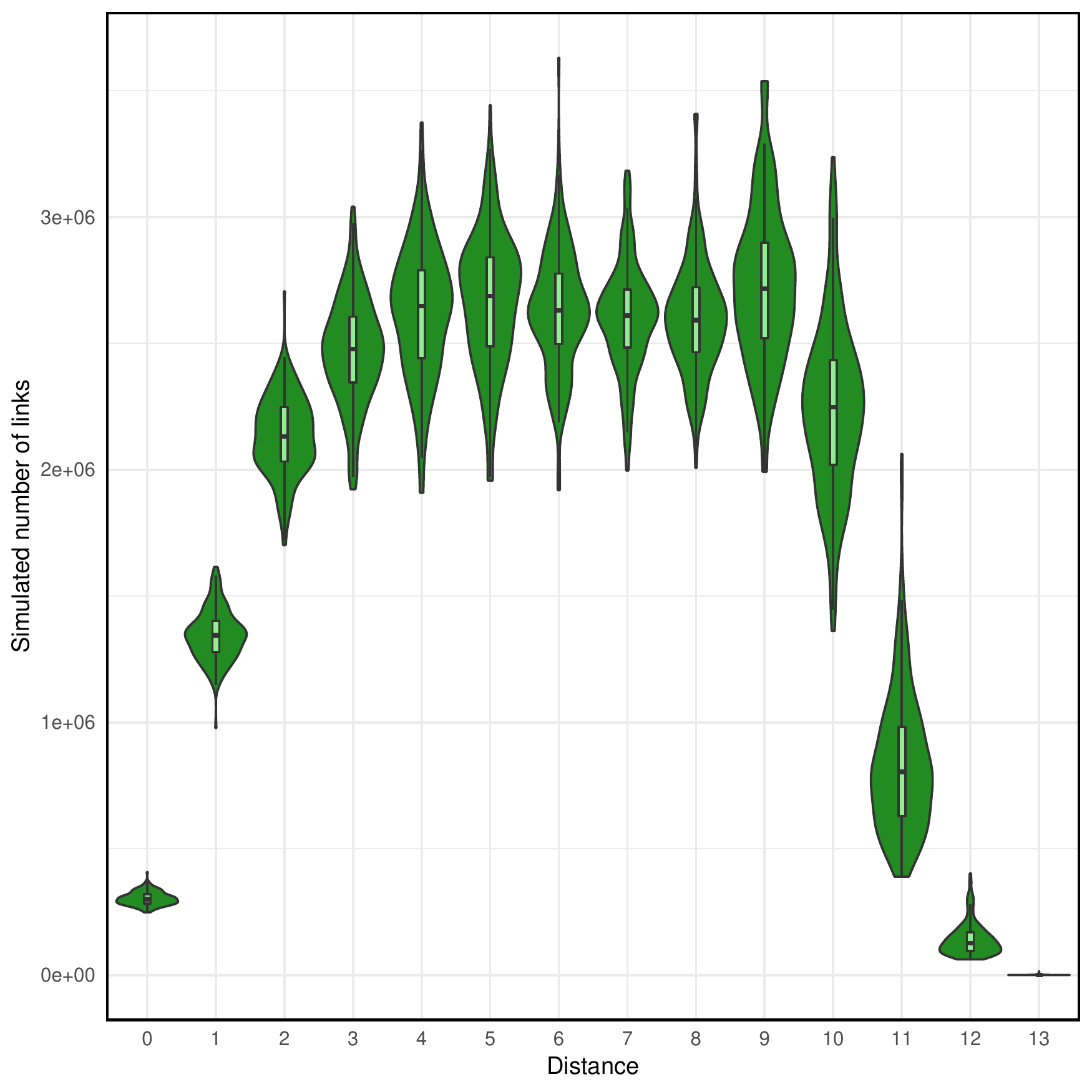} &
\includegraphics[width = 0.25\linewidth]{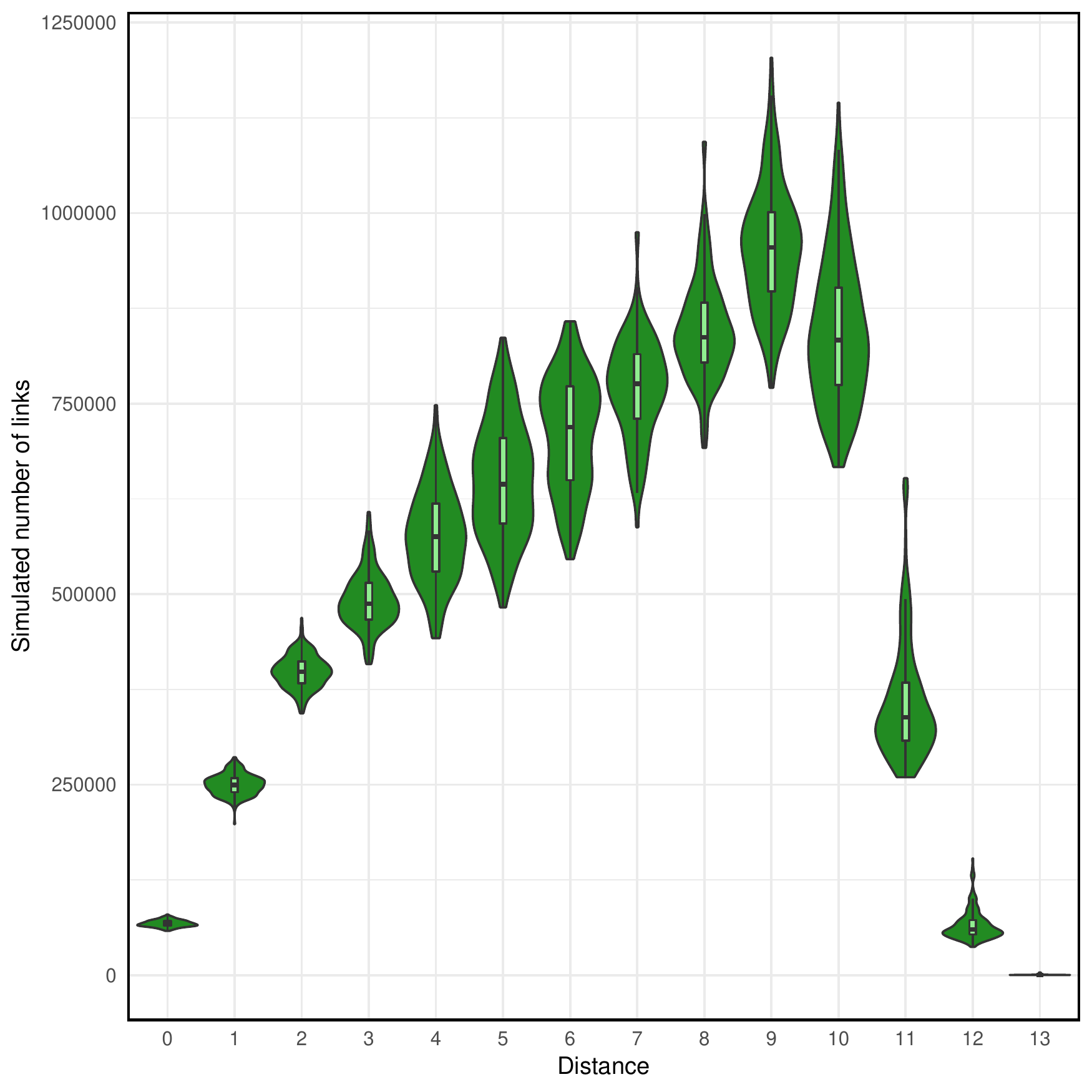}
\\
{\scriptsize following} & {\scriptsize starring} & {\scriptsize forking} \\
\includegraphics[width = 0.25\linewidth]{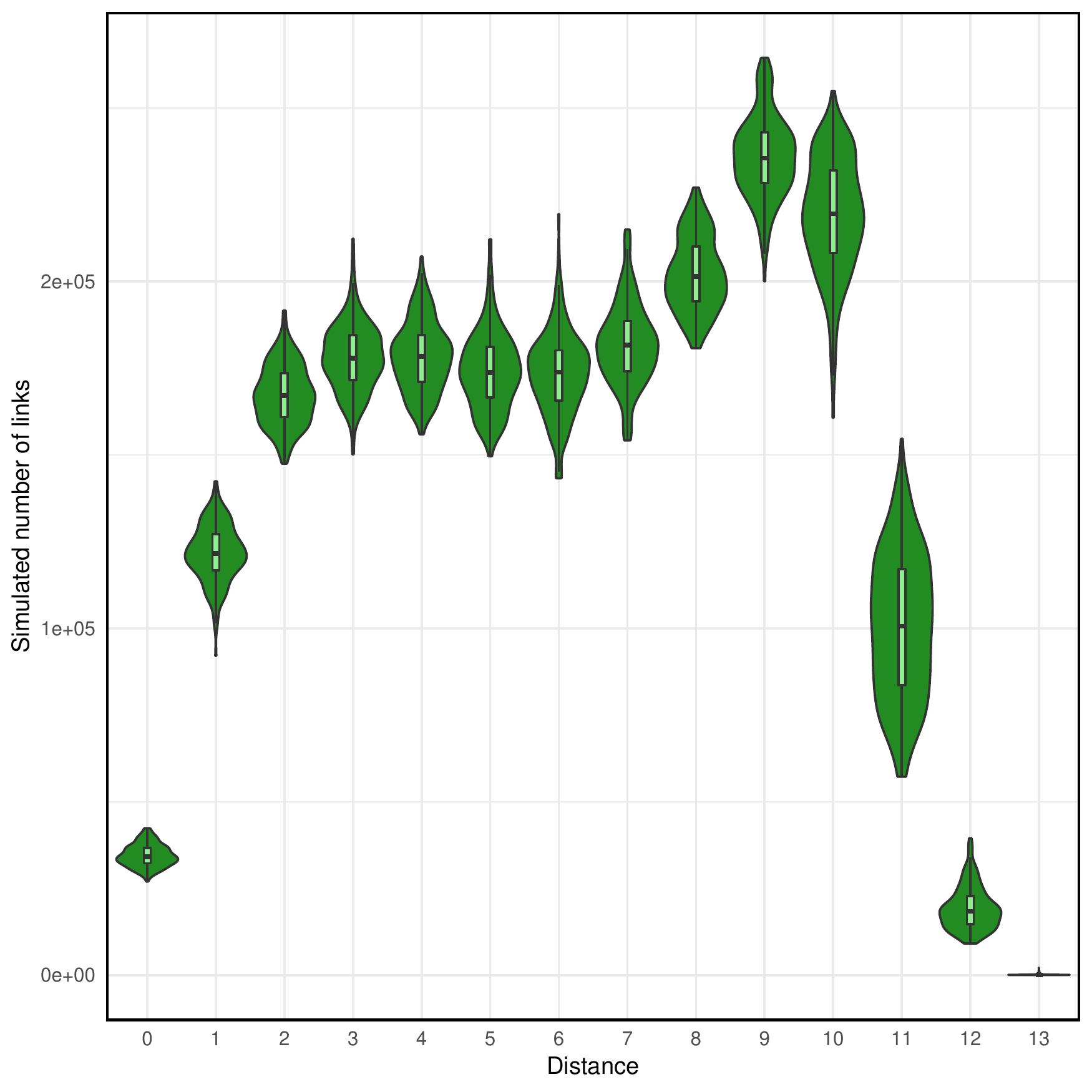} &
\includegraphics[width = 0.25\linewidth]{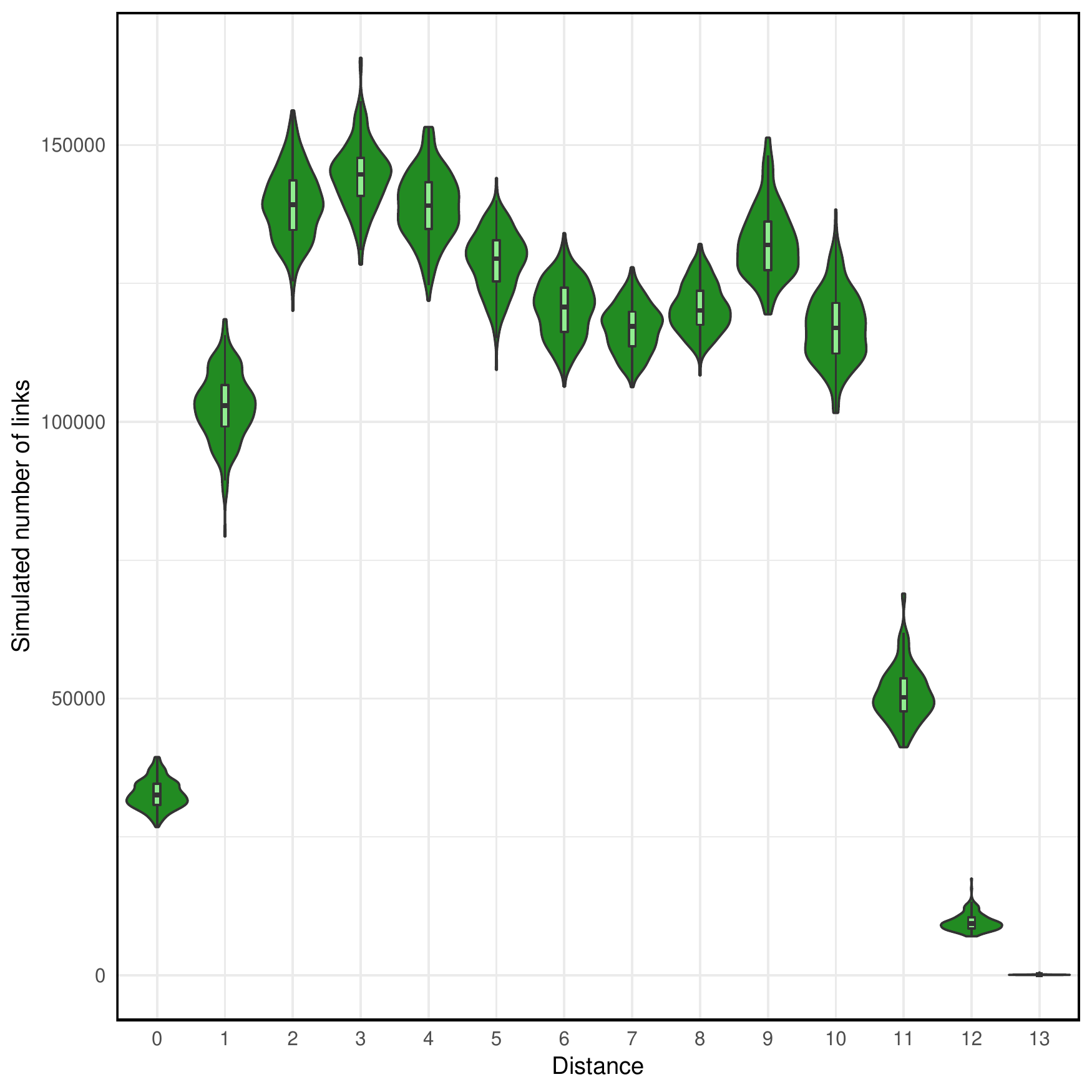} &
\includegraphics[width = 0.25\linewidth]{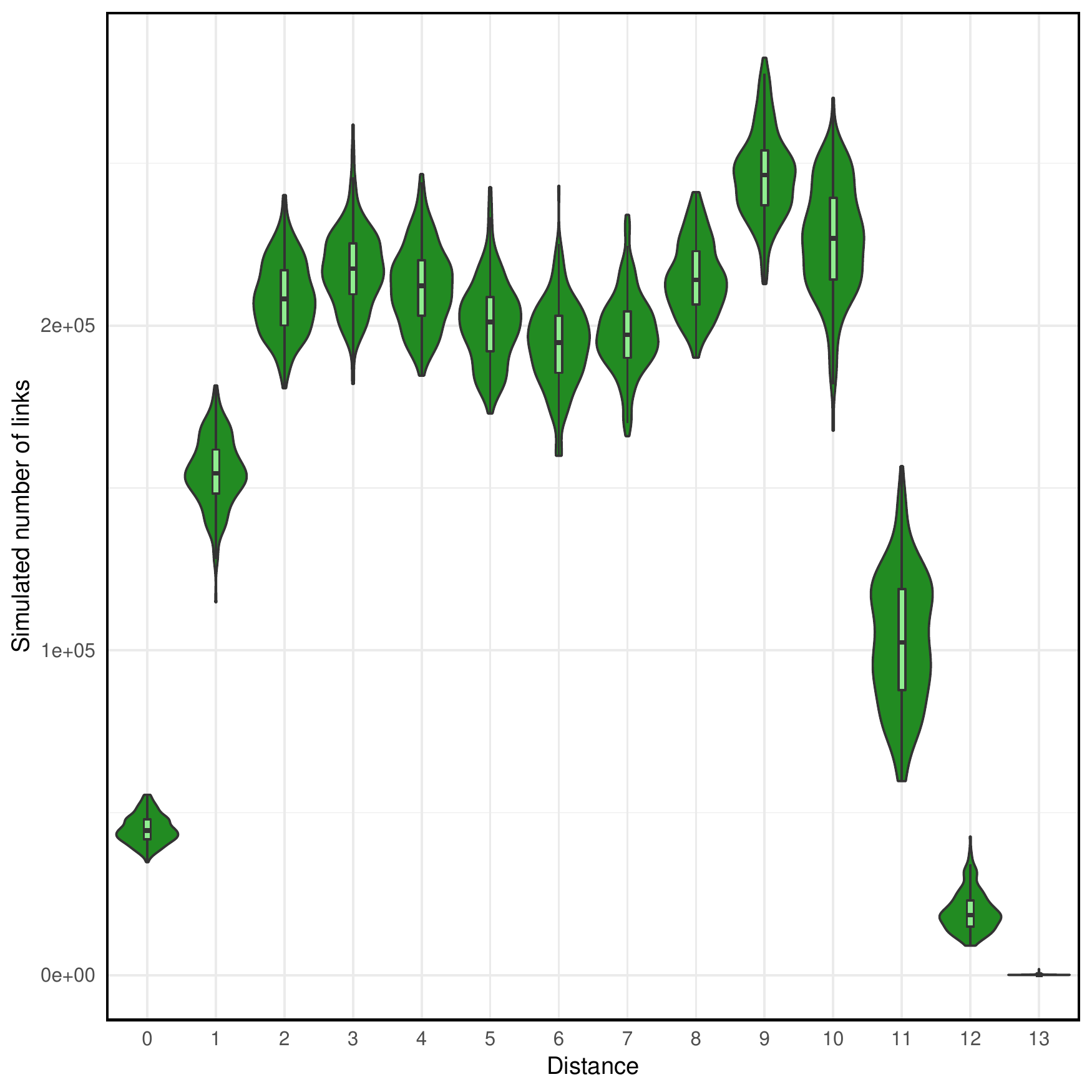} \\
{\scriptsize issues} & {\scriptsize pull requests} & {\scriptsize comments}
\end{tabular}
\caption{Distributions of the numbers of edges among developers in 200 simulations. The increasing distance (x axis) indicates increasing dissimilarity \label{kq-sims}}
\end{figure}

We aimed to find the patterns behind link creation. To this end, we counted how many edges in GitHub networks (following, starring, forking, issues, pulls, comments) occur between developers in a~given distance.
We notice that links between dissimilar users (very long distance) are rare \ref{kq-sims}. The shapes of the distributions of links differ based on the type of connection. We observe that distributions of the number of links in the networks related to information-flow or reputation mechanisms (e.g., following, starring) resemble bell-curves with a slight bias towards dissimilarity (long distances). The distributions in networks originating from the collaboration (issues, comments, pull requests) are two-modal. That suggests benefits from working with both similar and dissimilar users. Especially in the case of pull requests, we observe the bias towards the similarity. The distribution of the number of links for the forking network is intriguing. Forking is necessary for sending a pull request, so we would expect this network to behave like the collaboration-related ones. However, a significant number of forks is not intended for collaboration \cite{perils}. These forks serve as personal copies of the software, highlighting this network's reputation-building origin. It seems plausible to connect with and copy software of users who are dissimilar to us since it would be impossible or troublesome to produce this software by ourselves (e.g., due to the differences in coding experience). While both similarity and diversity play roles in collaboration, information flow and reputation lean towards diversity.  

\subsection{Role of diversity in GitHub teams}

We have already found that similarity favors collaboration on the source code in the form of pull requests.
However, following the literature suggestions on team-working, we suppose that the connections between similar and dissimilar individuals have diverse roles. We expect that similar users tend to work on the source code (lower barriers for communication, similar skills, or abilities). Simultaneously, diversity allows for reporting bugs or suggesting features (differently processed packages of information among heterogeneous individuals).

The different roles of communication between similar and diverse users emerge from the analysis of the messages' subjects. With greater similarity, the role of [user-mention] token increases; users talk about tasks and refer to others. On the contrary, communication from diverse users often refers to their experience or needs. The more diverse users, the more likely to talk about themselves than refer to others (for ``I/you'' fraction Spearman $\rho = 0.92$, p-value 0.00).

\paragraph{Development-related topics and diversity}

To analyze the roles of different clusters of links, we found the keywords in the distances.
We were interested in words related to collaboration on (modifying) code, bug reporting, asking for help, or suggesting features.
\footnote{The zoomable word clouds with frequent words used in the communication based on similarity are available here: \url{https://mimuw.edu.pl/~dot/en/pages/wordclouds.html} -- those visuals are mostly for Readers' entertainment, the paper contains crucial insights.}

In Table \ref{distances}, we present selected keywords for the mentioned groups. With the increasing dissimilarity of the users (a greater distance), the role of the token [codesnippet] and the words related to the writing the source code (e.g., pull request, merge) decreases; the role of the words related to bug reports and feature requests (e.g., error, problem, debug, help, howto) increases. Those relationships are statistically significant (joint significance tests).

\begin{table*}[!bt]
  \centering
\caption{Usage of common words related to project development in division by GitHub users' similarity. ***, **, * denote significance of  Spearman coefficient at 0.01, 0.05, 0.1 level, respectively.  \label{distances}}  
\resizebox{0.99\linewidth}{!}{
  \begin{tabular}{l|c|cc|ccccccccccc}
    \toprule
    group   & word            & \multicolumn{2}{c|}{correlation} & 0   & 1     & 2     & 3     & 4     & 5     & 6     & 7     & 8     & 9     & 10 \\
    \midrule
modify  & code-snippet    & -0.74                       & **  & 31.18 & 28.41 & 26.88 & 25.82 & 24.26 & 24.42 & 20.70 & 31.69 & 16.25 & 15.41 & 12.84 \\
        & pull-request    & 0.31                        &     & 0.91  & 1.29  & 1.46  & 1.70  & 1.89  & 1.76  & 3.20  & 5.37  & 1.87  & 1.59  & 1.26  \\
        & merge           & -0.77                       & *** & 3.73  & 3.57  & 3.69  & 3.67  & 3.59  & 3.85  & 3.24  & 2.46  & 2.73  & 3.27  & 1.43  \\
        & patch           & -0.55                       & *   & 1.08  & 1.05  & 1.13  & 1.16  & 1.11  & 1.09  & 1.07  & 0.77  & 0.85  & 0.86  & 1.08  \\ \hline
report  & error           & 0.86                        & *** & 3.20  & 3.77  & 4.27  & 4.65  & 4.85  & 5.16  & 6.30  & 5.99  & 6.43  & 7.69  & 10.04 \\
        & problem         & 0.81                        & *** & 4.22  & 4.79  & 5.24  & 5.63  & 5.68  & 5.85  & 5.94  & 4.74  & 6.31  & 7.72  & 9.97  \\
        & not-work        & 0.91                        & *** & 0.65  & 0.85  & 0.97  & 1.08  & 1.15  & 1.18  & 1.19  & 1.00  & 1.38  & 1.85  & 2.53  \\
        & bug             & 0.78                        & *** & 2.88  & 3.12  & 3.32  & 3.50  & 3.52  & 3.54  & 3.71  & 3.13  & 3.51  & 4.18  & 4.30  \\
        & bug-report      & 0.59                        & *   & 0.12  & 0.17  & 0.15  & 0.21  & 0.16  & 0.19  & 0.17  & 0.11  & 0.21  & 0.31  & 0.23  \\
        & failed          & 0.92                        & *** & 0.36  & 0.52  & 0.51  & 0.54  & 0.59  & 0.78  & 0.84  & 0.82  & 0.84  & 1.16  & 1.48  \\ \hline
help    & help            & 0.71                        & **  & 3.37  & 3.46  & 3.77  & 4.01  & 4.17  & 3.99  & 4.20  & 3.31  & 4.36  & 5.09  & 7.06  \\
        & howto           & 0.85                        & *** & 1.21  & 1.42  & 1.63  & 1.77  & 1.94  & 1.90  & 1.97  & 1.55  & 2.17  & 2.62  & 3.60  \\
        & not-know        & 0.54                        & *   & 0.82  & 1.02  & 1.11  & 1.14  & 1.15  & 1.12  & 1.12  & 0.85  & 1.10  & 1.24  & 1.48  \\ \hline
suggest & suggest         & 0.55                        & **  & 1.69  & 1.68  & 1.78  & 1.86  & 1.85  & 1.87  & 1.82  & 1.38  & 1.83  & 2.25  & 2.17  \\
& feature-request & 0.58                        & **  & 0.08  & 0.11  & 0.14  & 0.15  & 0.13  & 0.14  & 0.16  & 0.10  & 0.13  & 0.16  & 0.25  \\
\bottomrule 
  \end{tabular}}
\end{table*}

Interestingly, even if the correlation between talking about pull requests and the dissimilarity of the users is moderately positive, it is non-significant. A closer analysis of the artifacts containing this word written by the dissimilar users reveals that the comments are often reminders (e.g., \emph{I have sent you a pull request}), mentioning a user is new to GitHub. This may result from newcomers' unfamiliarity with the etiquette rules: they may perceive their (often the first) pull request as something extraordinary, instead of everyday activity for a developer. The lack of experience also correlates with the mistakes in pull requests (e.g., \emph{used wrong branch for pull request}, \emph{Added a pull request with the same issue description... sorry, new to GitHub!!} or \emph{please, disregard this pull request}). The comments also mention a lack of decision about the sent pull request or give the justification for closing it. Our observations are generally in line with the phenomenon described in \cite{tsay}. Pull requests that generate discussions are less likely to be accepted. 

Similar users are likely to talk about and analyze issues (for [github-issue] token Spearman $\rho = -0.87$, p-value 0.00) or point at other GitHub repositories as the reference (for [github-repo] token Spearman $\rho = -0.55$, p-value 0.008). On the contrary, with increasing diversity the proportion of automated messages increases ( Spearman $\rho = 0.53$, p-value 0.009). 

\paragraph{Learning from diversity}

According to Information Processing Theory, we may expect that increasing diversity of interlocutors
provides learning opportunities. Collaboration among newbies and experts \cite{allaho} internalizes different experiences and other intangible assets within the team. We have already found evidence for this phenomenon analyzing comments containing the word ``pull request''. Moreover, commenters dissimilar from the repository owner more often admit they are new to something (Spearman $\rho = 0.82$, p-value 0.002).

With increasing diversity, users are more likely to mention their willingness to learn something in the conversation (Spearman $\rho = 0.82$, p-value 0.003). Willingness to learn parallels with lack of experience (e.g., \emph{This is my first time doing this that's why I'm asking. I want to learn.}) or encountered errors (e.g., \emph{I want to learn because way I have done it in my site it cause late loading. How you done it?} [original wording]).

To combat the stereotype that developers tend to refrain from helping newbies by either referring to READMEs, manuals (as in Read The Manual), or publicly available resources (e.g., StackOverflow) we checked whether there are significant positive correlations with those words. While the correlations with StackOverflow or manual were positive, they were statistically insignificant (p-values greater than 0.20). On the contrary, the correlation with ``README'' was negative and statistically significant. The word README occurred more often among similar users. The comments indicated the need for updating it (e.g., \emph{Yeah README should be updated. thanks.} or \emph{we need more information in the README so the people would know how it works.}) or discussed what it should include (e.g., \emph{could we please list such dependencies in the README to make this more clear?} or \emph{would it beneficial to update the README given the problem I had with the explicit require?} [original wording]). 

\paragraph{Curator role of diversity}

Open Source is widely accepted as meritocracy \cite{vasilescu_1}, where decisions are made by deliberating groups \cite{tsay}. However, for multiple reasons, deliberating groups often converge on falsehood rather than truth \cite{hayek-challenge}. Deliberating groups of similar users are prone to biases.

The results on error reporting from Table \ref{distances} prove diversity's effectiveness in pooling dispersed information among GitHub users. Pull requests, even if accepted, may contain hidden errors. Users similar to the pull request author may not be able to find them. We get the solution to a problem (the source code is produced), but it is not optimal. In such a case, diversity serves as a medium that leads to optimal solutions by pointing out the errors and integrating different perspectives.

\subsection{Diversity and politeness}

Hardly can source code insult anyone. However, working on project development involves verbal communication and discussions over code-related decisions.
Recent studies have already started to investigate the role of politeness in collaborative software development \cite{ortu}.
Politeness can significantly impact the outcome of communication. We observe a statistically significant negative
correlation between the diversity of the interlocutors and mentioning others. There is no significant
connection between apologizing and diversity. However, with increasing dissimilarity, users are more probable to express gratitude and thank others (Spearman $\rho = 0.66$, p-value 0.03). In the usage of emoticons, greater diversity comes with more frowns than smileys (Spearman $\rho =: 0.90$, p-value 0.00). Similarity encourages users to upvote other's propositions (Spearman $\rho = -0.94$, p-value 0.00), which can be both positive and dangerous. On the one hand, upvoting shows admiration and warms up the conversation. On the other hand, it may foster echo chambers and lead to cognitive biases, limiting critical discourse possibilities.

The result concerning usage of emoticons and phatic communication (verbal or non-verbal communication of a~social function, e.g., starting a~conversation, greeting someone rather than an informative function) may suggest that the attitude of the interlocutors correlates with their similarity or not. To check this, we performed sentiment analysis at the level of sentences. Sentiment analysis is a~technique that identifies how sentiments are expressed in texts and whether they indicate positive or negative opinions \cite{sentiment}. We utilize the implementation of sentiment analysis in sentimentr R package \cite{sentimentr}. The dictionary was user-defined; it consisted of the combined and augmented Jockers \cite{jockers} \& Rinker's augmented Hu \& Liu \cite{huliu} positive/negative word list and the emoji/emoticon sentiment lexicon \cite{emoji}.

\begin{figure*}[bt]
\centering
\includegraphics[width=\linewidth]{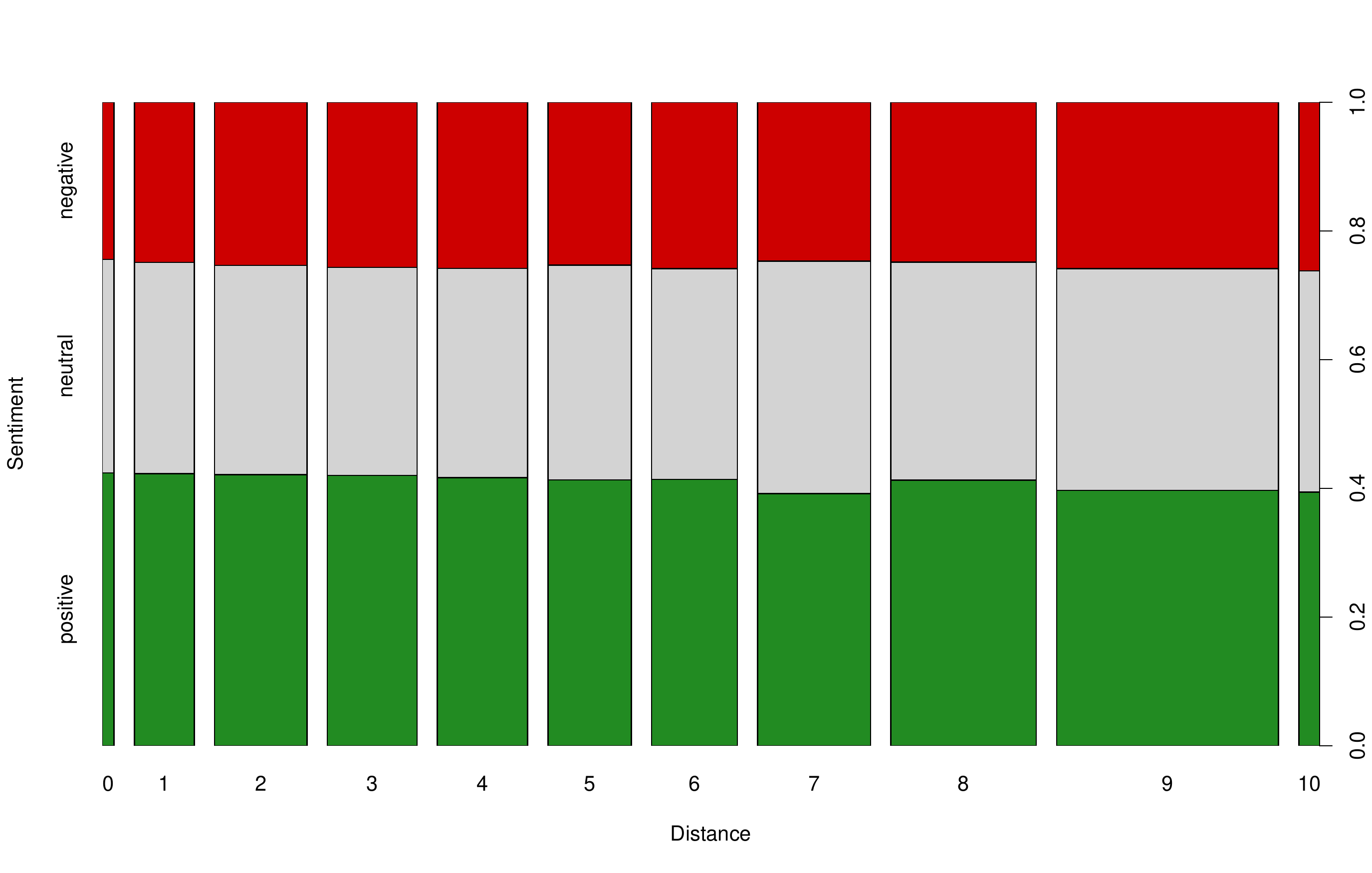}
\caption{Sentiment distribution in communication among GitHub users of increasing diversity \label{sentiment}}
\end{figure*}

Based on Figure \ref{sentiment}, we notice that the proportions among the sentiments are similar: there are more positive (green) or neutral (grey) comments than the negative ones (red). Polarity and similarity are not independent ($\chi^2$ = 11866, p-value 0.000). Spearman $\rho =-0.01$, p-value 0.000, indicates there is a~weak negative correlation between the distance and the polarity. The increasing dissimilarity of users correlates with more negativity in discussions.

\section{Robustness of the results}\label{sec:diag}

Utilizing a mixed-methods approach always comes with questions about the quality of its output. This section discusses factors that could affect the obtained results and how we mitigated them.

\paragraph{Choosing the manifold for SOM}
One of the major concerns regarding SOM design is the choice of the manifold. We checked if Klein quartic choice could bias the conclusions. We compared the results obtained from: Euclidean disk, Hyperbolic disk, Sphere, Elliptic Plane, Torus, minimal quotient, genus 2 quotient, genus 14 quotient. Results for Klein bottle were indistinguishable from torus (p-values of the Kolmogorov-Smirnov tests always higher than 0.05), and the Macbeath surface (genus 7 quotient) needed a significantly higher number of neurons to implement, thus was incomparable%
\footnote{The interactive versions of the results of the experiments with SOM's underlying topology are available here: \url{https://mimuw.edu.pl/~dot/resources/wizualizacje/github-rogueviz/ghweb.html}  -- an extended technical meta-analysis is out of the scope of this paper, we described the steps of robustness checks.}.%
We adapted the standard model-selection criteria:

\begin{enumerate}
\item Log-likelihood $\ell_d$ -- SOM distances' goodness in predicting links between developers
\item Log-likelihood $\ell_c$ -- clustering's goodness in predicting links between developers
\item Akaike Information Criterion (AIC) based on clustering -- a~standard tool used in assessing the relative quality of statistical models for a~given data set
\item Bayesian Information Criterion (BIC) based on clustering -- AIC could underestimate the role of the number of neurons, so we added a~second tool assessing the relative quality of statistical models for a~given data set.
\end{enumerate} 

%\begin{table}[!bt]
%  \centering
%  \caption{Basic information on the manifolds chosen for robustness checks \label{manifolds_data}}
%\begin{tabular}{lcc} \toprule
%  Manifold          & Maximum Distance & Number of neurons \\ \midrule
%  Euclidean disk    & 26 & 520 \\ 
%  Hyperbolic disk   & 14 & 520 \\ 
%  Sphere            & 20 & 522 \\ 
%  Elliptic plane    & 15 & 541 \\ 
%  Torus             & 16 & 520 \\ 
%  Minimal quotient  & 17 & 524 \\ 
%  Genus 2 quotient  & 16 & 520 \\ 
%  Genus 3 quotient  & 13 & 528 \\
%  Genus 14 quotient & 9  & 520 \\ \bottomrule
%\end{tabular}
%\end{table} 

Hyperbolic quotient spaces: minimal quotient, genus 2 quotient, genus 3 quotient (Klein quartic), and genus 14 quotient had satisfactory predictive properties. The clustering based on Klein quartic also returns a~relatively low number of distances and provides aesthetically pleasing results due to high symmetry. That is why we chose it.  

\paragraph{Randomness in clustering} The strength of the SOM roots within their ability to learn the patterns in
an unsupervised way. We start by randomizing weights for neurons, which we may perceive as a~source
of randomness. One could argue that the results are not stable and significantly depend on the initial weights \cite{soms}. To prevent this, we conducted simulations. For every manifold, we run the algorithm independently 200 times. Means and confidence intervals were computed using leave-one-out simulations.

\paragraph{Comparison with random networks} One could argue if the resulting shapes of the distributions of the number of links in the given distance stemmed from the degree distribution of those networks. To check this, we simulated the number of links in the given distance against the random networks with the algorithm based on the Markov chain Monte Carlo method. %
%In each iteration, we drew pairs of edges from the given network and swapped the end nodes between them. If both new edges did not contain self-loops and did not exist in the network, we replaced the original edges with the swapped ones. After $100$ times number of edges  iterations we obtained a~randomized network, preserving the degree distribution.
We conducted 100 such simulations and afterward checked the equality of distributions with the Kolmogorov-Smirnov test. In all cases, we rejected the null hypothesis.

\paragraph{Comparison with other methods} We decided to construct a~similarity measure based on non-Euclidean SOMs. However, other approaches are possible. SOMs are generally preferred over PCA because they provide better insights when data is complex, diverse or the relationships among variables are non-linear \cite{koh-pca}. The results of PCA also lack the elegant spatial interpretation. However, it could be arguable whether we benefit from the information generated by the SOM's underlying topology. To check this, we benchmarked the results of SOMs against the naive method. We determined the distribution of the values of the Euclidean distance between the observations by simulation and then grouped those values to define the distances. The values of log-likelihoods $\ell_d$ for distances defined with the naive method were worse than any value of $\ell_d$ for non-Euclidean SOM. 

\section{Discussion and implications}
This paper presented an analysis of the nature of GitHub users' connections and the role of users' diversity during project development. Our sample consisted of both big and small projects, experts and newbies. Whereas, as Cosentino et al. \cite{cosentino} state, most papers on GitHub use data sets of small-medium size, generated with poor sampling techniques, we performed big data analysis. Robustness checks proved our results are generalizable.

\subsection{Homophily and heterophily among GitHub users}
To answer RQ1, we conducted a one-dimensional and multidimensional analysis. The one-dimensional analysis allowed us
to investigate homophily or heterophily from the point of links' costs. The multidimensional analysis provided us insights on the quality of the relationships. To conveniently discuss the results, we will divide the networks into two groups: code-related networks (issues, comments, pull requests) and reputation-related networks (following, starring, forking).  

Every action performed within the network bears a~cost. Even if the information is a~valuable good, the process of obtaining data can be expensive. The costs are of different kinds: being it the alternative cost of time spent on gathering data, the possibility of overflooding with information ending with the disability to find the relevant one, or the cost of maintaining connections with the information sources.

In GitHub, the flow of information mostly occurs within the following network: one is given easily accessible notifications about the activity of the entities of interest. However, the following is expensive: one should strategically choose the links which will provide valuable information. If one establishes contact with a~node influential in more than one network, they reduce the cost of obtaining information: connection with this one node provides information from more than one network.

The overall heterophily observed for reputation-related networks is unsurprising. In the literature, popular and influential users of GitHub are known as rockstars. A~rockstar has a~large number of followers interested in their coding activity, what projects they follow or work on \cite{herbsleb, dabbish, badashian}. The number of followers serves as a~signal of popularity/status in the community. Popular users are loosely interconnected among them -- usually have more connections with unpopular users than with other popular ones \cite{lima, allaho}. We observe the same phenomenon while analyzing rich-clubs and assortativity measures concerning the directionality.

The cost of maintaining a~connection varies among the networks. Following, starring or forking networks are characterized by a~relatively low cost of link starting and maintaining compared to collaboration activities with high entry costs, especially sending a~pull request. Maintaining multiple connections usually requires some optimization; either we limit the number of connections or the number of nodes we keep relationships with. Choosing an entity that is influential in more than one network at the same time enables directly relates to such optimization; the agents may also perceive multiple links with a~given entity more valuable than the occasional, disposable connection.

The results for the pull request network are consistent with the similarity attraction theory. Technical collaboration on the source code is easier for similar individuals. They have a lower cost of communication.

The number of connections provides little information about their quality. We assume that neglecting the impact of the quality and the strength of the connections distorts the networks' analysis significantly. Our analysis shows that insights drawn from the number of links are not necessarily the same when considering the implied quality of the relations.

The networks we analyze in this study are characterized by power-law, which means that the creation of the connection may be driven by up to a~certain degree non-random factors, e.g., reputation or preferential attachment. The impact of creating an additional link is non-linear. In such a~situation, two randomly chosen links may bear different costs and value. Ideally, one could also investigate the strength of the connections by utilizing weighted graphs; in this study, we present a simplified approach. Establishing more valuable connections also relates to the optimization issues: it may reduce at the same time the number of nodes and the connections one maintains.

Even if by average GitHub users are driven by heterophily in the number of connections, analysis of the code-related characteristic in a multidimensional way reveals that the links' nature is not that straightforward. While significant bias towards heterophily in forking network is explained by creating a copy of the software -- it is more convenient to make one instead of writing code by ourselves -- heterophily in other reputation-generated networks is not as much visible.

We observe that connections among very similar and very diverse users are rare. The explanation hides exactly in the definition of reputation. Reputation is about differences: we need them to construct hierarchy. For the practical purposes of information gathering, it is beneficial to connect with users who are ``different but not too much different'' from the given user.
Too much similarity reduces the possibilities for finding something new because of the substitution in skills (or even echo-chambering); too much diversity may require skills that are not within one's reach. On the contrary, collaboration-oriented networks benefit from both homophily and heterophily. While homophily fosters code production in the form of pull requests (as in similarity attraction theory), heterophily allows for knowledge dissemination and finding solutions due to differently processed packages of information (as in information processing theory).

\subsection{Roles of the links}
To answer RQ2, we conducted a text mining analysis. Our results are consistent with information processing theory: a diverse team, consisting of more and less experienced users, aggregates different perspectives. Our findings on collaboration between newbies and experts, and the learning possibilities arising from such team-building support Allaho et al. \cite{allaho} suggestion about the environmental characteristics of GitHub co-working.

It is not easy to compare our results to other research on the impact of diversity on collaboration in GitHub due to notable methodological differences. First, we analyzed more facets of collaboration than only productivity measured in source code. Second, other researchers analyzed the impact of socio-demographic factors (often inferred, e.g., gender \cite{vasilescu_1, ortu} or nationality \cite{ortu}), and we decided to construct a generalized similarity measure based on development-related, observable characteristics. Future research direction could involve checking if the roles of diverse links differ based on the size of the project users contribute to.

\subsection{The sentiment of conversation}
To answer RQ3, we conducted a sentiment analysis. Our results support similarity attraction theory and SIC that suggest that work-group heterogeneity can lead to confusion, stress, and conflict.

Our results of sentiment analysis revealed that, on average, GitHub users are polite and respectful. The correlation between the negativity in sentiment and increasing dissimilarity of interlocutors although significant is much weaker than the correlations observed for the positive aspects of the team's diversity.

\subsection{Implications}
Groups can be prone to amplification of errors or judgment (e.g., reputational) biases \cite{hayek-challenge}. Such groups can solve the problem (as in the definition), but the result is not optimal. We stress the curator role of diversity. In GitHub, homogeneity fosters code production. However, even accepted pull request might contain errors, hard to notice in the team of similar skills. Diversity encourages bug reports and improvement suggestions, coordinating to optimal solutions. We find it a social implication of grave importance.

Previous research focused on the roles of given characteristics (e.g., gender, tenure) on project success, usually equal to the amount of code. That, in turn, may lead to the impression that contribution to Open Source projects is impossible without technical skills. We highlight that successful collaboration refers not only to the source code. We show that non-technicians (non-developers) have their place in the Open Source teams – they are essential for the project development. Low programming skills should not be the reason to refrain from participation. One does not have to code to help the project evolve. Non-developers help with improvements or mundane but necessary tasks (e.g., bug reporting).

To practitioners, even if the project managers may have incentives to look for code developers, we show it is beneficial to attract diverse collaborators. Focusing on commits encourages to follow policies that discriminate non-technicians while constructing a team. Sure, we get the code produced, but do not control for its quality. The curator role of diversity gives a rationale for welcoming users who were afraid of joining, e.g., due to the stereotype ``coding is crucial''.

\section{Threats to validity}

\paragraph{Lack of qualitative information} Data mining can unintentionally be misused and can then produce results that appear to be significant but may stem from the actual sample and cannot be reproduced later. Therefore we are cautious with the results we obtained: we aim to find the hidden patterns behind the creation of the links among users without asking those users what the actual reasons for establishing links are. This limitation is hard to mitigate: in our opinion, in recent years, surveying developers was significantly overused and led to the reluctance of the respondents. Furthermore, even if we carried out the survey, the common concerns related to survey data would apply, e.g., regarding the sampling techniques, the generalizability of the results, and even the declarative responses' validity.

\paragraph{Lack of the thread structure} We lack information on the sequence of the comments in issues threads, which limits the possibilities for the text analysis. Our sample does not contain the responses of the owners of the repositories; we have only part of the discourse. We are not able to conduct directional analysis, e.g., check if the keywords in the responses or their polarity differ between similar and dissimilar users. Lack of the structure also hinders possibilities for temporal analysis of the communication. In future work, we are going to mitigate these limitations.

\paragraph{Limiting discussions to English} One could argue that the analysis of the communication in English is not representative for the GitHub users. We do not find this limitation threatening -- English is the most widely used language among GitHub users (our rough estimation on the available data returns over 85\% of the content).
English is also correctly detected with CLD2. The cases of misclassified language as English were rare (a few occurrences). In contrast, misclassifying English as a~different language was more common due to the specificity of the IT-related content and frequent occurrences of proper names. E.g., short comments containing ``github'' were usually classified as Danish, and short descriptions containing ``emacs'' as Siswant.

Emojis serve as a~non-verbal means of communication. The comments written entirely with them were the second most common in the hierarchy (over 500,000 artifacts). Even if it is arguable whether Emoji is a~language \cite{emojis-lang}, we included them in the analysis. Comments containing nothing more than an emoticon/emoji were extracted with regular expressions and labeled as ``Emoji''. Because of the lack of Chinese or Spanish knowledge we did not perform the analysis for those languages. The usage of other languages was negligible.

\section{Conclusions}
Remaining mindful of various confounds, we claim the following contributions.
\begin{itemize}
\item With statistical modeling and large-scale analysis, we investigated the tendencies behind link creation among GitHub users. We found that diversity plays a~crucial role in links' creation among users who exchange information (e.g., in issues, comments, and following networks). On the contrary, similar users establish connections in networks related to actual coding. 
\item Our quantitative analysis was augmented by text mining on the discussions in GitHub. This way, we found that teams' diversity encourages bug-reporting and user experience (reporting problems, suggesting facilities). Diverse users have opportunities to learn from each other via collaborating.
\end{itemize}  

We collected a data set on millions of users and repositories, merging available data sources on GitHub. Our data set contained captured or inferred information on users' characteristics, their connections in six social networks, and the textual artifacts generated by them. Our approach significantly advantages over the previous survey-based, small-sample research. We use high-quality data; the information we process is not declarative, so the agents lack incentives to cheat. To our best knowledge, this is the first study of GitHub users to combine Social Network Analysis of more than two networks, classic statistical interference, and qualitative study of textual artifacts.

On average, developers are driven by heterophily, which can be explained by the results derived from team collaboration analyses in fields different from software development. Having a~diversified team leads to diversified knowledge assets, making it easier to find optimal solutions. In the case of teams made of homogeneous agents, the only advantage resulting from adding another entity (assuming that we can easily divide tasks into smaller and smaller parts) is that the work is done in a shorter time. None of them can propose a~solution that cannot be derived by other members. However, working in a~diversified team can be more complicated than working in a~more or less homogeneous team -- there are some communication costs. That is why we also observe some sort of homophily among Open Source developers collaborating via GitHub. 

Disassortativity usually leads to the accumulation of different information packages within the team, so the group's ability to collectively solve various problems increases. The disassortative mixing in GitHub networks shows that the collaboration occurs among newbies and experts (as suggested in \cite{allaho}), presenting the possibilities of internalizing different experiences and other intangible assets within the team. We provide empirical evidence for the collective intelligence phenomena in GitHub. 

This study refers to unresolved concerns about teamwork. Groups can be prone to amplification of errors or judgment (e.g., reputational) biases \cite{hayek-challenge}. Such groups can solve the problem (as in the definition), but the result is not optimal. We stress the curator role of diversity. In GitHub, homogeneity fosters code production. Even accepted pull requests might contain errors, hard to notice in the team of similar skills. Diversity encourages bug reports and improvement suggestions, coordinating to optimal solutions.

We contribute to both academic research as well as industry practice. To practitioners, our results provide insight into what characteristics they should pay attention to while seeking new collaborators.

This work was supported by the National Science Centre, Poland, grant DEC-2016/21/N/HS4/02100.

\bibliographystyle{alpha}

\begin{thebibliography}{VKHSW12}

\bibitem[AE12]{soms}
Umut Asan and Secil Ercan.
\newblock An introduction to self-organizing maps.
\newblock In C.~Kahraman, editor, {\em Computational Intelligence Systems in
  Industrial Engineering: with Recent Theory and Applications}, pages 299--319.
  Atlantis Press, 2012.

\bibitem[AL13]{allaho}
Mohammad~Y. Allaho and Wang-Chien Lee.
\newblock Analyzing the social ties and structure of contributors in open
  source software community.
\newblock In {\em 2013 IEEE/ACM International Conference on Advances in Social
  Networks Analysis and Mining (ASONAM 2013)}, pages 56--60, 2013.

\bibitem[Als16]{emojis-lang}
Hamza Alshenqeeti.
\newblock Are emojis creating a new or old visual language for new generations?
  a socio-semiotic study.
\newblock {\em Advances in Language and Literary Studies}, 7(6):56--69, 2016.

\bibitem[BDE14]{ergms}
Michael~J. Bannister, William~E. Devanny, and David Eppstein.
\newblock Ergms are hard.
\newblock Technical report, 2014.

\bibitem[BL01]{bergquist}
Magnus Bergquist and Jan Ljungberg.
\newblock The power of gifts: organizing social relationships in open source
  communities.
\newblock {\em Information Systems Journal}, 11(4):305--320, 2001.

\bibitem[BR06]{brossi2006}
Andrea Bonaccorsi and Cristina Rossi.
\newblock Comparing motivations of individual programmers and firms to take
  part in the open source movement: From community to business.
\newblock {\em Knowledge, Technology {\&} Policy}, 18(4):40--64, 2006.

\bibitem[BS16]{badashian}
Ali~Sajedi Badashian and Eleni Stroulia.
\newblock Measuring user influence in github: The million follower fallacy.
\newblock In {\em Proceedings of the 3rd International Workshop on
  CrowdSourcing in Software Engineering}, CSI-SE '16, pages 15--21, 2016.

\bibitem[Bur05]{burt}
Ronald~S. Burt.
\newblock {\em Brokerage and Closure. An introduction to Social Capital}.
\newblock Oxford University Press, 2005.

\bibitem[Byr71]{byrne}
Erwin Byrne, Donn.
\newblock {\em The attraction paradigm}.
\newblock Personality and psychopathology. Academic Press, 1971.

\bibitem[CCIC17]{cosentino}
Valerio Cosentino, Javier~L. Cánovas~Izquierdo, and Jordi Cabot.
\newblock A systematic mapping study of software development with github.
\newblock {\em IEEE Access}, 5:7173--7192, 2017.

\bibitem[Cel16]{celinska_euromed}
Dorota Celi\'nska.
\newblock {Information and influence in social network of Open Source
  community}.
\newblock In {\em 9th Annual Conference of the EuroMed Academy of Business},
  pages 485--495, 2016.

\bibitem[Cel18]{celinska_smes}
Dorota Celi\'nska.
\newblock {Coding together in a social network: collaboration among GitHub
  users}.
\newblock In {\em Proceedings of the 9th International Conference on Social
  Media and Society}, pages 31--40, 2018.

\bibitem[CFSV06]{colizza}
Vittoria Colizza, Alessandro Flammini, M.~Angeles Serrano, and Alessandro
  Vespignani.
\newblock Detecting rich-club ordering in complex networks.
\newblock {\em Nature Phisics}, 2:110--115, 2006.

\bibitem[CLRS01]{cormen}
Thomas~H. Cormen, Charles~E. Leiserson, Ronald~R. Rivest, and Clifford Stein.
\newblock {\em Introduction to Algorithms}.
\newblock The MIT Press, 2001.

\bibitem[CS15]{chatziasimidis}
Fragkiskos Chatziasimidis and Ioannis Stamelos.
\newblock Data collection and analysis of github repositories and users.
\newblock In {\em 2015 6th International Conference on Information,
  Intelligence, Systems and Applications (IISA)}, pages 1--6, 2015.

\bibitem[CWHW08]{crowston}
Kevin Crowston, Kangning Wei, James Howison, and Andrea Wiggins.
\newblock Free/libre open-source software development: What we know and what we
  do not know.
\newblock {\em ACM Computing Surveys}, 44(2):7:1--7:35, 2008.

\bibitem[DAS13]{daniel}
Sherae Daniel, Ritu Agarwal, and Katherine Stewart.
\newblock The effects of diversity in global, distributed collectives: A study
  of open source project success.
\newblock {\em Information Systems Research}, 24:312--333, 06 2013.

\bibitem[DCG16]{koh-pca}
Gopal Das, Manojit Chattopadhyay, and Sumeet Gupta.
\newblock A comparison of self-organising maps and principal components
  analysis.
\newblock {\em International Journal of Market Research}, 58(6):815--834, 2016.

\bibitem[DS08]{david_shapiro}
Paul~A. David and Joseph~S. Shapiro.
\newblock Community-baased production of open-source software: What do we know
  about the developers who participate?
\newblock {\em Information Economics and Policy}, 1(20):364--398, 2008.

\bibitem[DSTH12]{dabbish}
Laura Dabbish, Colleen Stuart, Jason Tsay, and Jim Herbsleb.
\newblock Social coding in github: Transparency and collaboration in an open
  software repository.
\newblock In {\em Proceedings of the ACM 2012 Conference on Computer Supported
  Cooperative Work}, CSCW '12, pages 1277--1286, 2012.

\bibitem[FGBM14]{onboarding}
Fabian Fagerholm, Alejandro Guinea, Jay Borenstein, and Jürgen Münch.
\newblock Onboarding in open source projects.
\newblock {\em IEEE Software}, 11 2014.

\bibitem[For69]{fortes}
Meyer Fortes.
\newblock {\em Kinship and the Social Order: the Legacy of Lewis Henry Morgan}.
\newblock Transaction Publishers, 1969.

\bibitem[Gou]{ght2}
Georgios Gousios.
\newblock {GHTorrent}.
\newblock \url{https://www.ghtorrent.org/}.
\newblock Online: Dec. 5, 2017.

\bibitem[Gri12]{gha}
Ilya Grigorik.
\newblock {Github Archive}.
\newblock \url{https://www.githubarchive.org/}, 2012.
\newblock Online: Dec. 5, 2017.

\bibitem[GS12]{ght_firehose}
Georgios Gousios and Diomidis Spinellis.
\newblock Ghtorrent: Github's data from a firehose.
\newblock In {\em 9th {IEEE} Working Conference of Mining Software
  Repositories, {MSR} 2012, June 2--3, 2012, Zurich, Switzerland}, pages
  12--21, 2012.

\bibitem[Hey99]{heylighen}
Francis Heylighen.
\newblock Collective intelligence and its implementation on the web: Algorithms
  to develop a collective mental map.
\newblock {\em Computational {\&} Mathematical Organization Theory},
  5(3):253--280, 1999.

\bibitem[HH07]{horwitz}
Sujin~K. Horwitz and Irwin~B. Horwitz.
\newblock The effects of team diversity on team outcomes: A meta-analytic
  review of team demography.
\newblock {\em Journal of Management}, 33(6):987--1015, 2007.

\bibitem[HL04]{huliu}
Minqing Hu and Bing Liu.
\newblock Mining and summarizing customer reviews.
\newblock In {\em Proceedings of the Tenth ACM SIGKDD International Conference
  on Knowledge Discovery and Data Mining}, KDD '04, pages 168--177, New York,
  NY, USA, 2004. ACM.

\bibitem[HNH03]{hertel}
Guido Hertel, Sven Niedner, and Stefanie Herrmann.
\newblock Motivation of software developers in open source projects: an
  internet-based survey of contributors to the linux kernel.
\newblock {\em Research Policy}, 32(7):1159--1177, 2003.

\bibitem[HO02]{hars}
Alexander Hars and Shaosong Ou.
\newblock Working for free? motivations for participating in open-source
  projects.
\newblock {\em International Journal of Electronic Commerce}, 6(3):25--39,
  2002.

\bibitem[HRS04]{hann}
Il-Horn Hann, Jeff Roberts, and Sandra Slaughter.
\newblock Why developers participate in open source software projects: an
  empirical investigation.
\newblock {\em International Conference on Information Systems 2004}, 2004.

\bibitem[HTSD13]{herbsleb}
James Herbsleb, Jason Tsay, Coleen Stuart, and Laura Dabbish.
\newblock Leveraging transparency.
\newblock {\em IEEE Software}, 30:37--43, 2013.

\bibitem[JGJ{\etalchar{+}}14]{jarczyk}
Oskar Jarczyk, B{\l}a{\.{z}}ej Gruszka, Szymon Jaroszewicz, Leszek Bukowski,
  and Adam Wierzbicki.
\newblock Github projects. quality analysis of open-source software.
\newblock In Luca~Maria Aiello and Daniel McFarland, editors, {\em Social
  Informatics: 6th International Conference, SocInfo 2014, Barcelona, Spain,
  November 11--13, 2014. Proceedings}, pages 80--94, Cham, 2014. Springer
  International Publishing.

\bibitem[Joc17]{jockers}
Matthew~L. Jockers.
\newblock Syuzhet: Extract sentiment and plot arcs from text, 2017.
\newblock version 1.0.4.

\bibitem[KGB{\etalchar{+}}14]{perils}
Eirini Kalliamvakou, Georgios Gousios, Kelly Blincoe, Leif Singer, Daniel~M.
  German, and Daniela Damian.
\newblock The promises and perils of mining github (extended version).
\newblock In {\em Proceedings of the 11th Working Conference on Mining Software
  Repositories}, MSR 2014, pages 92--101. ACM, 2014.

\bibitem[Koh97]{kohonen}
Teuvo Kohonen, editor.
\newblock {\em Self-organizing Maps}.
\newblock Springer-Verlag, Berlin, Heidelberg, 1997.

\bibitem[Lev97]{levy}
Pierre Levy.
\newblock {\em Collective Intelligence: Mankind's Emerging World in
  Cyberspace}.
\newblock Perseus Books, Cambridge, MA, USA, 1997.

\bibitem[LH13]{litvak}
Nelly Litvak and Remco Hofstad.
\newblock Uncovering disassortativity in large scale-free networks.
\newblock {\em Physical review. E, Statistical, nonlinear, and soft matter
  physics}, 87(2):1--11, 2013.

\bibitem[Lin01]{lin}
N.~Lin.
\newblock {\em Social Capital: A Theory of Social Structure and Action}.
\newblock Cambridge University Press, 2001.

\bibitem[LM54]{lazarsfeld}
Paul~Felix Lazarsfeld and Robert~K. Merton.
\newblock Friendship as a social process: a substantive and methodological
  analysis.
\newblock In P.L.~Kendall P.F.~Lazarsfeld, editor, {\em The Varied Sociology}.
  Columbia University Press, 1954.

\bibitem[Lou08]{louadi}
Mohamed Louadi.
\newblock Knowledge heterogeneity and social network analysis - towards
  conceptual and measurement clarifications.
\newblock {\em Knowledge Management Research \& Practice}, 6:199--213, 09 2008.

\bibitem[LRM14]{lima}
Antonio Lima, Luca Rossi, and Mirco Musolesi.
\newblock Coding together at scale: Github as a collaborative social network.
\newblock {\em Eighth International AAAI Conference on Weblogs and Social
  Media}, 2014.

\bibitem[LS05]{latteman}
Christoph Lattemann and Stefan Stieglitz.
\newblock Framework for governance in open source communities.
\newblock In {\em Proceedings of the 38th Annual Hawaii International
  Conference on System Sciences}, pages 192a--192a, 2005.

\bibitem[LT02]{lerner}
Josh Lerner and Jean Tirole.
\newblock Some simple economics of open source.
\newblock {\em Journal of Industrial Economics}, 50:197--234, 2002.

\bibitem[LvH03]{lakhani_hippel}
Karim~R. Lakhani and Eric von Hippel.
\newblock How open source software works: "free" user-to-user assistance.
\newblock {\em Research Policy}, 32(6):923--943, 2003.

\bibitem[LW05]{lakhani}
Karim Lakhani and Robert Wolf.
\newblock Why hackers do what they do: Understanding motivation and effort in
  free/open source software projects.
\newblock In Joeseph Feller, Brian Fitzgerald, and Hissam Scott, editors, {\em
  Perspectives on Free and Open Source Software}. MIT Press, Cambridge, 2005.

\bibitem[Mar88]{marsden}
Peter~V. Marsden.
\newblock Homogeneity in confiding relations.
\newblock {\em Social Networks}, 10(1):57 -- 76, 1988.

\bibitem[MB12]{meissonier}
Regis Meissonier and Isabelle Bourdon.
\newblock Toward an enacted approach to understanding oss developers
  motivations.
\newblock {\em International Journal of Technology and Human Interactions},
  8(2):38--54, 2012.

\bibitem[MBPG14]{mcdonald}
Nora McDonald, Kelly Blincoe, Eva Petakovic, and Sean Goggins.
\newblock Modelling distributed collaboration on github.
\newblock {\em Advances in Complex Systems}, 17(07n08):14500--14524, 2014.

\bibitem[MDH13]{marlow}
Jennifer Marlow, Laura Dabbish, and Jim Herbsleb.
\newblock Impression formation in online peers production: Activity traces and
  personal profiles in github.
\newblock In {\em Proceedings of 2013 Conference on Computer Supported
  Cooperative Work}, pages 117--128, 2013.

\bibitem[MSLC01]{birds}
Miller McPherson, Lynn Smith-Lovin, and James~M Cook.
\newblock Birds of a feather: Homophily in social networks.
\newblock {\em Annual Review of Sociology}, 27(1):415--444, 2001.

\bibitem[New02]{newman}
Mark E.~J. Newman.
\newblock Assortative mixing in networks.
\newblock {\em Physical Review Letters}, 89(20):208701, October 2002.

\bibitem[NSSM15]{emoji}
Kralj~Petra Novak, Jasmina Smailović, Borut Sluban, and Igor Mozetič.
\newblock Sentiment of emojis.
\newblock {\em PLOS ONE}, 10(12):1--22, 12 2015.

\bibitem[NVM15]{noldus}
Rogier Noldus and Piet Van~Mieghem.
\newblock Assortativity in complex networks.
\newblock {\em Journal of Complex Networks}, 3(4):507--542, 03 2015.

\bibitem[NY03]{sentiment}
Tetsuya Nasukawa and Jeonghee Yi.
\newblock Sentiment analysis: Capturing favorability using natural language
  processing.
\newblock In {\em Proceedings of the 2Nd International Conference on Knowledge
  Capture}, K-CAP '03, pages 70--77, New York, NY, USA, 2003. ACM.

\bibitem[ODC{\etalchar{+}}17]{ortu}
Marco Ortu, Giuseppe Destefanis, Steve Counsell, Stephen Swift, Roberto
  Tonelli, and Michele Marchesi.
\newblock How diverse is your team? investigating gender and nationality
  diversity in github teams.
\newblock {\em Journal of Software Engineering Research and Development},
  5(1):9--27, 2017.

\bibitem[OR07]{osterloh}
Margit Osterloh and Sandra Rota.
\newblock Open source software development--just another case of collective
  invention?
\newblock {\em Research Policy}, 36(2):157--171, 2007.

\bibitem[Ray01]{raymond2}
Eric~S. Raymond.
\newblock {\em The Cathedral and the Bazaar: Musings on Linux and Open Source
  by an Accidental Revolutionary}.
\newblock O'Reilly \& Associates, Inc., Sebastopol, CA, USA, 2001.

\bibitem[RHS06]{roberts}
Jeffrey~A. Roberts, Il-Horn Hann, and Sandra~A. Slaughter.
\newblock Understanding the motivations, participation, and performance of open
  source software developers: A longitudinal study of the apache projects.
\newblock {\em Management Science}, 52(7):984--999, 2006.

\bibitem[Rin18]{sentimentr}
Tyler~W. Rinker.
\newblock {sentimentr}: Calculate text polarity sentiment, 2018.
\newblock version 2.6.1.

\bibitem[Sal12]{salminen}
Juho Salminen.
\newblock Collective intelligence in humans: A literature review.
\newblock {\em ArXiv}, abs/1204.3401, 2012.

\bibitem[Sit13]{cld2}
Richard~L. Sites.
\newblock Compact language detector 2, 2013.

\bibitem[SP78]{salancik}
Gerald~R. Salancik and J.~Alan Pfeffer.
\newblock A social information processing approach to job attitudes and task
  design.
\newblock {\em Administrative science quarterly}, 23 2:224--53, 1978.

\bibitem[Sun06]{hayek-challenge}
Cass~R. Sunstein.
\newblock Deliberating groups versus prediction markets (or hayek's challenge
  to habermas).
\newblock {\em Episteme}, 3(3):192–213, 2006.

\bibitem[Sur05]{surowiecki}
James Surowiecki.
\newblock {\em The Wisdom of Crowds}.
\newblock Anchor, 2005.

\bibitem[Taj82]{tajfel}
Henri Tajfel.
\newblock Social psychology of intergroup relations.
\newblock {\em Annual Review of Psychology}, 33(1):1--39, 1982.

\bibitem[TDH14]{tsay}
Jason Tsay, Laura Dabbish, and James Herbsleb.
\newblock Let’s talk about it: Evaluating contributions through discussion in
  github.
\newblock In {\em Proceedings of the 22nd ACM SIGSOFT International Symposium
  on Foundations of Software Engineering}, FSE 2014, page 144–154, New York,
  NY, USA, 2014. Association for Computing Machinery.

\bibitem[TKM{\etalchar{+}}17]{gender}
Josh Terrell, Andrew Kofink, Justin Middleton, Clarissa Rainear, Emerson
  Murphy-Hill, Chris Parnin, and Jon Stallings.
\newblock Gender differences and bias in open source: pull request acceptance
  of women versus men.
\newblock {\em PeerJ Computer Science}, 3, 2017.

\bibitem[VFS15]{vasilescu_2}
Bogdan Vasilescu, Vladimir Filkov, and Alexander Serebrenik.
\newblock Perceptions of diversity on git hub: {A} user survey.
\newblock In {\em CHASE@ICSE}, pages 50--56. {IEEE} Computer Society, 2015.

\bibitem[VKHSW12]{vonkrogh}
Georg Von~Krogh, Stefan Haefliger, Sebastian Spaeth, and Martin~W. Wallin.
\newblock Carrots and rainbows: Motivation and social practice in open source
  software development.
\newblock {\em MIS Quarterly}, 36(2):649--676, June 2012.

\bibitem[VPR{\etalchar{+}}15]{vasilescu_1}
Bogdan Vasilescu, Daryl Posnett, Baishakhi Ray, Mark G.~J. van~den Brand,
  Alexander Serebrenik, Premkumar~T. Devanbu, and Vladimir Filkov.
\newblock Gender and tenure diversity in github teams.
\newblock In {\em {CHI}}, pages 3789--3798. {ACM}, 2015.

\bibitem[Wan08]{github-blog-11}
Chris Wanstrath.
\newblock {GitHub: Free for Open Source}.
\newblock \url{https://github.com/blog/11}, 2008.
\newblock Online: Aug. 6, 2017.

\bibitem[Zei03]{zeitlyn}
David Zeitlyn.
\newblock Gift economies in the development of open source software:
  anthropological reflections.
\newblock {\em Research Policy}, 32(7):1287--1291, 2003.

\bibitem[ZM04]{zhou_richclub}
Shi Zhou and Raul~J. Mondragon.
\newblock The rich-club phenomenon in the internet topology.
\newblock {\em Communications Letters, IEEE}, 8:180--182, 2004.

\end{thebibliography}
\newcommand{\etalchar}[1]{$^{#1}$}

\end{document}